\documentclass[aps,prc,superscriptaddress,twoside,twocolumn,nofootinbib,showpacs]{revtex4-2}

\usepackage{amsmath,amssymb}
\usepackage{graphicx}
\usepackage{braket}
\usepackage{slashed}

\usepackage{newtxtext,newtxmath}
\usepackage{mathrsfs}
\usepackage{csquotes}
\usepackage{dcolumn,tensor,physics}

\usepackage{hyperref}
\hypersetup{colorlinks=true,citecolor=blue,linkcolor=blue,urlcolor=blue}
\newcommand{\hrefr}[2]{\href{#2}{#1}}

\allowdisplaybreaks

\begin{document}


\title{\boldmath Analysis of virtual meson production in solvable (1+1) dimensional scalar field theory}

\author{Yongwoo Choi}
\email{sunctchoi@gmail.com}
\affiliation{Department of Physics, Kyungpook National University, Daegu 41566, Korea}

\author{Ho-Meoyng Choi}
\email{homyoung@knu.ac.kr}
\affiliation{Department of Physics Education, Teachers College, Kyungpook National University, Daegu 41566, Korea}

\author{Chueng-Ryong Ji}
\email{ji@ncsu.edu}
\affiliation{Department of Physics, North Carolina State University, Raleigh, NC 27695-8202, USA}

\author{Yongseok Oh}
\email{yohphy@knu.ac.kr}
\affiliation{Department of Physics, Kyungpook National University, Daegu 41566, Korea}
\affiliation{Asia Pacific Center for Theoretical Physics, Pohang, Gyeongbuk 37673, Korea}


\begin{abstract}
Light-front time-ordered amplitudes are investigated in the virtual scalar meson production process in $(1+1)$ 
dimensions using the solvable scalar field theory extended from the conventional Wick-Cutkosky model. 
There is only one Compton form factor (CFF) in the $(1+1)$ dimensional computation of the virtual meson 
production process, and we compute both the real and imaginary parts of the CFF for the entire 
kinematic regions of $Q^2>0$ and $t<0$.
We then analyze the contribution of each and every light-front time-ordered amplitude to the CFF as a function of 
$Q^2$ and $t$. 
In particular, we discuss the significance of the ``cat's ears'' contributions for gauge invariance and 
the validity of the ``handbag dominance'' in the formulation of the generalized parton distribution (GPD) 
function used typically in the analysis of deeply virtual meson production processes. 
We explicitly derive the GPD from the ``handbag'' light-front time-ordered amplitudes in the $-t/Q^2 \ll 1$
limit and verify that the integrations of the GPD over the light-front longitudinal momentum fraction 
for the DGLAP and ERBL regions correspond to the valence and nonvalence contributions of the 
electromagnetic form factor that we have recently reported [Phys. Rev. D \textbf{103}, 076002 (2021)].
We also discuss the correspondence of the GPD to the parton distribution function for the analysis of 
the deep inelastic lepton-hadron scattering process and the utility of the new light-front longitudinal
spatial variable $\tilde{z}$.
\end{abstract}


\maketitle


\section{Introduction}  
\label{sec:1}

One of the main goals in hadron physics is to understand the properties and structure of hadrons in terms of 
quarks and gluons.
Since the elastic electron scattering from the nucleon unveiled the non-pointlike structure of the 
nucleon~\cite{HM55}, the information on the spatial distributions of charge and current inside the nucleon 
has been obtained from the electromagnetic (EM) form factors. 
The information on the momentum distributions of quarks and gluons inside nucleons can also be accessed by 
the parton distribution functions (PDFs) measured through deep inelastic scattering (DIS) 
processes~\cite{BFKB69}.

Since the DIS with the longitudinally polarized beam and the polarized target proton yielded the surprising 
result that the quarks and antiquarks inside the proton carry about only 30\% of the total spin of 
the proton~\cite{EMC89}, one of the key observables to explore the so-called `proton spin puzzle' has 
been the set of generalized parton distributions 
(GPDs)~\cite{Ji96a,Ji96b,Radyushkin96a,Radyushkin97b,MRGDH94,GPV01,Diehl03,BR05,CJK01}.
GPDs describe internal parton structures of the hadron, unifying the investigation of form factors and PDFs.  
In particular, the parton's orbital angular momentum contribution to the nucleon spin may be estimated 
with the formulation of GPDs. 
The extraction of GPDs from the experimental data can be accessed mainly by hard exclusive electroproduction
processes such as the deeply virtual Compton scattering~(DVCS) and the deeply virtual meson production
(DVMP) processes. 
In these processes, an electron interacts with a parton from the hadron, e.g., the nucleon, by the exchange 
of a virtual photon, and the struck parton radiates a real photon (DVCS process) or hadronizes into a meson 
(DVMP process)~\cite{FGHK15,Tarrach75,Metz97,BM08,KM09,BJ14}.
Both in DVCS and DVMP, the GPD formalism relies on the ``handbag dominance'' representing the factorization 
of the hard and soft parts in the respective scattering amplitudes. 
Here, the light-front dynamics (LFD) plays an important role in providing both the skewness $\zeta$ and the
light-front (LF) longitudinal momentum fraction $x$ of the parton struck by the probing virtual 
photon off the target. It is well known that the integrals of the leading-twist GPDs in the $s$- and 
$u$-channel handbag amplitudes of both DVCS and DVMP processes carry the factorized denominator factors 
such as $1/(x-\zeta)$ and $1/x$, respectively.

While the virtual Compton scattering (VCS) process is coherent with the Bethe-Heitler (BH) process~\cite{BH34}, 
the virtual meson production (VMP) process does not possess complications from the involvement of the BH
process and offers a unique way for experimental exploration of the hadronic structure for the study of 
Quantum Chromodynamics and strong interactions. 
In particular, the coherent electroproduction of pseudoscalar ($J^{PC}=0^{-+}$) or scalar ($0^{++}$) mesons off 
a scalar target (for example, the \nuclide[4]{He} nucleus~\cite{CLAS17b}) provides an excellent experimental terrain
to discuss the fundamental nature of hadrons without involving much complication from the spin degrees of 
freedom.
In Ref.~\cite{JCLB18}, two of us discussed the most general formulation of the differential cross sections 
for the meson ($0^{-+}$ or $0^{++}$) production processes which involve only one or two hadronic form factors,
respectively, when the target is a scalar particle. 
In particular, the beam spin asymmetry was discussed and our findings from the general formulation were 
contrasted with respect to the GPD formulation.

In the present article, we investigate the electroproduction process of a scalar meson off a scalar target, simulating
for example, $\gamma^{*} + \nuclide[4]{He} \to f_{0}(980) + \nuclide[4]{He}$, in the one-loop level of an exactly solvable (1+1) dimensional scalar field theory extended from the conventional Wick-Cutkosky model~\cite{Wick-Cutkosky}. 
The same scalar field model theory was previously applied to the analysis of the longitudinal charge density~\cite{CCJO21}.
As the transverse rotations are absent in $(1+1)$ dimensions, the advantage of the LFD with the LF time 
$x^+ = x^0 + x^3$ as the evolution parameter is maximized in contrast to the usual instant form dynamics (IFD) 
with the ordinary time $x^0$ as the evolution parameter. 
In LFD, the individual $x^+$-ordered amplitudes contributing to the hadronic form factor are invariant 
under the boost, i.e., frame-independent, while the individual $x^0$-ordered amplitudes in IFD are not 
invariant under the boost but dependent on the reference frame. 
As only one hadronic form factor is involved in ($1+1$) dimensions for the electroproduction process of a scalar meson 
off a scalar target, the analysis may be regarded relatively simple without involving the beam spin asymmetry. 
In this work, we focus on analyzing the essential features of the LFD by benchmarking the $(1+1)$ dimensional 
characteristics of the VMP process.

The extraction of the hadronic form factor, the so-called Compton form factor (CFF), is made by utilizing
the general formulation of the hadronic currents presented in Ref.~\cite{JCLB18}.
The real and imaginary parts of the CFF are extracted explicitly. 
In order to explore the applicability of the handbag dominance adopted in the GPD formulation, we extract 
the GPD in the DVMP limit and verify that the integrations of the GPD over $x$ for the DGLAP ($1> x> \zeta$) 
and ERBL ($\zeta > x > 0$) regions correspond to the respective valence and nonvalence contributions of the 
electromagnetic form factor that we have recently presented in Ref.~\cite{CCJO21}. 
The correspondence of the GPD to the PDF is also discussed with the new LF longitudinal 
spatial variable $\tilde z = x^- p^+$ recently introduced in Ref.~\cite{MB19}.  
We then contrast the CFF obtained from the GPD formulation with the CFF result from the general formulation 
of the VMP process in the present exactly solvable scalar field model.

This article is organized as follows. 
In Section~\ref{sec:2}, we present the kinematics of the virtual meson production process off the scalar target.
Section~\ref{sec:3} is devoted to the derivation of the exact form of the CFF in the VMP process within the one-loop level of the 
scalar field model in $(1 + 1)$ dimensions. 
Complete analyses for various LF time-ordered diagrams involved in the VMP process are 
presented as well.
In Section~\ref{sec:4}, we extract the GPD, PDF, longitudinal probability density (LPD) in the LF coordinate 
space, and the EM form factor in the DVMP limit.
Section~\ref{sec:5} presents our numerical results for the CFF, GPD, PDF, LPD, and EM form factor of the 
scalar target simulating the mass arrangement of the $\gamma^{*} + \nuclide[4]{He} \to f_{0}(980) + \nuclide[4]{He}$ process.
We summarize and conclude in Section~\ref{sec:6}.

\section{Kinematics} 
\label{sec:2}

We begin with the kinematics involved in the virtual-photon scattering off the scalar target ($\mathcal{M}$) for the production of the scalar meson ($S$),
\begin{equation}
\label{eq1c}
\gamma^{*}(q) +\mathcal{M}(p) \rightarrow S(q')+\mathcal{M}(p'),
\end{equation}
where the initial (final) scalar target state is characterized by the momentum $p~(p')$ and the incoming 
virtual-photon and the outgoing meson by $q$ and $q'$, respectively. 
We shall use the component notation $a=(a^+,a^-)$ in $(1+1)$ dimensions and the metric is specified by 
$a^{\pm}=a^0\pm a^3$ and $a\cdot b=(a^+ b^- + a^- b^+)/2$.

Defining the four momentum transfer $\Delta = p - p'$, we have
\begin{eqnarray} 
\label{eq2c}
p &=& \left( p^{+}, ~~  \frac{M_{T}^{2}}{p^{+}} \right),\nonumber\\
p' &=& \left( (1-\zeta) p^{+},  ~~ \frac{1}{p^{+}} \Big(M_{T}^{2} - \frac{t}{\zeta}\Big) \right),
\end{eqnarray}
and
\begin{equation}
\label{eq3c}
\Delta = \Big(~\zeta p^{+},~~\frac{t}{\zeta p^{+}}~\Big),
\end{equation} 
where $M_T$ is the target mass and $\zeta =\Delta^+/p^+$ is the skewness parameter describing the asymmetry 
in plus momentum. 
The squared momentum transfer then reads 
\begin{equation}
\label{eq4c}
t=\Delta^2 = 2 p\cdot\Delta = -\zeta^2 M^2_T/(1-\zeta) \leq 0,
\end{equation}
which defines $\zeta$ in terms of $t$ as
\begin{equation} 
\label{eq5c}
\zeta  = \frac{1}{2 M_{T}^{2}} \left( t+\sqrt{t^{2} - 4 t M_{T}^{2}} \right),
\end{equation}
so that $0\leq \zeta\leq 1$ is taken. 
Considering the fact that $q^+\neq 0$ in $(1+1)$ dimensions, we choose the momenta $q$ and $q'$ as
\begin{eqnarray}   
\label{eq6c}
q &=& \left( \left( \mu_{s} \zeta'-\zeta \right) p^{+}, ~~ \frac{Q^2}{p^{+}} \Big( \frac{1}{\zeta'} 
+ \frac{\tau}{\zeta} \Big) \right), \nonumber\\
q' &=& \Big(~\mu_{s} \zeta' p^{+},~~ \frac{Q^2}{\zeta'~ p^{+}}~\Big),
\end{eqnarray}
where $ \mu_{s} = M_{S}^2/Q^2$ and $\tau=-t/Q^{2}$ with $q^2=-Q^2$ and $M_S$ being the mass of the produced 
scalar meson. 
It also gives the definition of $\zeta'$ through the relation with $\zeta$ as 
\begin{equation}   
\label{eq7c}
\frac{\zeta'}{\zeta} = \frac{2}{1+\mu_{s}-\tau+\sqrt{(1+\mu_{s}+\tau)^{2}-4\tau}},
\end{equation}
where $0 \leq \mu_{s} \zeta' < \zeta \leq 1$ is taken in deriving Eq.~(\ref{eq7c}) so that $q^+ <0$.
The Bjorken variable $x_{\rm Bj}^{} = Q^{2}/(2 p\cdot q)$ is then given by
\begin{equation}   
\label{eq8c}
x_{\rm Bj}^{} = \frac{2t}{t(1+\mu_s +\tau) - \sqrt{t(t- 4 M^2_T) [(1+\mu_s + \tau)^2 - 4\tau]}}.
\end{equation}
The maximum value of $x_{\rm Bj}^{}$ for a given value of $Q^2$ is obtained by the condition that
\begin{equation}  
\label{eq9c}
\left. \frac{d x_{\rm Bj}}{dt} \right|_{t=t_{\rm th}} = 0,
\end{equation} 
which determines the threshold momentum transfer squared $t_{\rm th}(Q^2)$ as  
\begin{equation}  
\label{eq10c}
t_{\rm th}(Q^{2}) = -\frac{M_{T}^{} \left( M_{S}^{2}+Q^{2} \right)}{ M_{T}+M_{S} }.
\end{equation} 
This corresponds to the threshold point where momentum directions of the outgoing target and the produced 
meson are swapped in the center-of-momentum frame.

In deeply virtual limit where $Q^2$ is very large compared to other scales of $M^2_T$, $M^2_S$, and $\abs{t}$, 
one can easily find that $\zeta \simeq x_{\rm Bj}$, i.e., $\zeta$ plays the role of $x_{\rm Bj}$.
We also note that $\zeta$ and $t$ are not independent in $(1+1)$ dimensions
while they are in general independent of each other in $(3+1)$ dimensions because of the nonzero transverse 
component of $\Delta$.
Explicitly, we have $t=- (\zeta^2 M^2_T + \Delta^2_\perp)/(1-\zeta)$ in $(3+1)$ dimensions~\cite{CJK01} where
$\Delta_\perp^2$ is the transverse momentum transfer squared. 
Defining the skewness parameters $\zeta=\zeta_{1+1}$ and $\zeta=\zeta_{3+1}$ in $(1+1)$ and $(3+1)$ 
dimensions, respectively, one can obtain the allowed range of $\zeta_{3+1}$ as $0 \leq \zeta_{3+1} \leq \zeta_{1+1}$ 
for a fixed value of $t$. 
Here, we note that the limit $t \to 0$ implies $\zeta_{1+1} \to 0$, while there is no such correlation between 
$t \to 0$ and $\zeta_{3+1} \to 0$ unless $\Delta_\perp^2 \to 0$ is imposed as well. 
This indicates that the ($1+1$) dimensional computation simulates only the forward production of the meson 
in the (3+1) dimensional computations as expected intuitively. 
As shown in Eq.~(\ref{eq4c}), the value of $-t$ is also not independent of the target mass $M_T$ in the ($1+1$) 
dimensional computations. 
Thus, for a given $-t$ value, the skewness parameter $\zeta_{1+1}$ gets smaller as $M_T$ increases. 
The consequence of such constraint in ($1+1$) dimensions will be revealed in the comparison of the CFF 
between the predictions from the GPD formulation deduced in the DVMP limit and our exact VMP computations 
obtained in the present work. 
In particular, the condition that $Q^2 \gg M^2_T$ may not be required for the DVMP limit in the $(1+1)$ dimensional 
computations because of the correlation among $\zeta_{1+1}$, $t$, and $M^2_T$. 
This would then imply that the condition $Q^2 \gg M^2_T$ may also not be required for the forward production
of mesons in the (3+1) dimensional computations.

\section{Model Calculations for virtual Meson Production} 
\label{sec:3}

In general, the total scattering amplitude $\mathscr{M}^{\mu}_{ \rm tot}$ for scalar meson production off
the scalar target of Eq.~(\ref{eq1c}) is expressed in terms of two independent CFFs~\cite{JCLB18} as%
\footnote{The two CFFs given by Eq.~(13) in~Ref.~\cite{JCLB18} are obtained from the replacement of
$\mathcal{F}_1 \to -({F}_1 + {F}_2)$ and $\mathcal{F}_2\to -{F}_2$ in Eq.~(\ref{eq11c}). }
\begin{eqnarray}
\label{eq11c}
\mathscr{M}^{\mu}_{\rm tot} &=& \left[ \left( \Delta \cdot q \right) q^{\mu}- q^2 \Delta^{\mu} \right]  
\mathcal{F}_{1}
+ \left[ \left( \Delta \cdot q \right) \mathcal{P}^{\mu} - \left( \mathcal{P} \cdot q \right) \Delta^{\mu} \right] 
\mathcal{F}_{2}
\nonumber\\
&\equiv& A^{\mu} \mathcal{F}_{1}+B^{\mu} \mathcal{F}_{2},
\end{eqnarray}
which defines $A^\mu$ and $B^\mu$, where $\mathcal{P}=p+p'$. 
The EM current conservation in Eq.~(\ref{eq11c}) is assured by the condition $q \cdot \mathscr{M}_{\rm tot}=0$.
The CFFs are measurable physical quantities and are related to the GPDs in the deeply virtual kinematic region, 
e.g., $Q^2 \gg -t$.
For the VCS process, it is not possible to distinguish whether the emitted real photon comes 
from the loop process in the hadronic sector or from the scattered electron, i.e., the BH process~\cite{BH34}. 
However, the VMP does not have a BH-type process since the scalar meson cannot be emitted from the electron.

Furthermore,  the two CFFs $\mathcal{F}_1$ and $\mathcal{F}_2$ in Eq.~(\ref{eq11c}) are not linearly 
independent in ($1+1$) dimensions since the two covariant vectors $A^{\mu}$ and $B^{\mu}$ are parallel to 
each other, i.e., $A^\mu = c B^\mu$, where the scaling factor $c$ reads
\begin{equation}
\label{eq12c}
c=2~\sqrt{\frac{t (t-4 M_{T}^{2})}{M_{S}^{4}+2 M_{S}^{2}(Q^{2}-t)+(Q^{2}+t)^{2}}},
\end{equation} 
for $\mu = \pm$. 
This leads us to redefine $\mathscr{M}^{\mu}_{\rm tot}$ in $(1+1)$ dimensions with one CFF as 
\begin{equation}
\label{eq13c}
\mathscr{M}^{\mu (1+1)}_{\rm tot} = \left[ \left( \Delta\cdot q \right) q^{\mu} - q^2 \Delta^{\mu} \right] 
\mathcal{F},
\end{equation}
where $\mathcal{F} = \mathcal{F}_1 + c \mathcal{F}_2$.

In the following, we shall perform the LF calculations of $\mathscr{M}^{\mu (1+1)}_{\rm tot}$ using the 
exactly solvable model based on the covariant Bethe-Salpeter (BS) calculations of $(1+1)$-dimensional 
scalar field theory. 
As the beam spin asymmetry for scalar meson production is absent in ($1+1$) dimensions because of the 
singleness of CFF, we do not need to involve the beam helicity but just focus on the LF calculations of the 
covariant scalar field model. 
We analyze the detailed structure of scattering amplitude coming from the loop diagrams below.

\subsection{Amplitudes from loop diagrams} 
\label{sec:3-A}

\begin{figure}[t!]\centering
\includegraphics[width=\columnwidth]{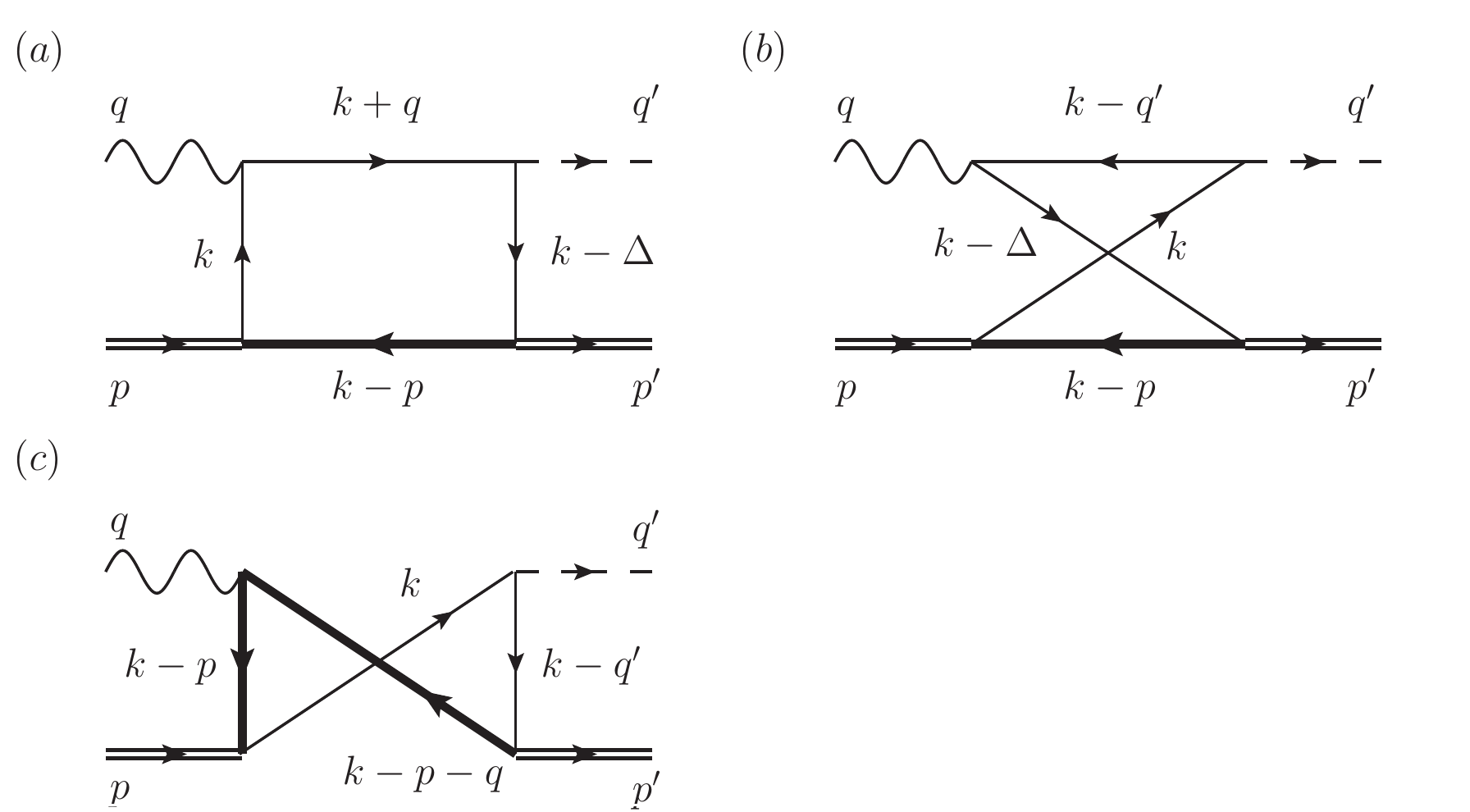}
\caption{\label{fig1}%
Relevant Feynman diagrams for the reaction of $\gamma^{*}(q) +\mathcal{M}(p) \to S(q')+\mathcal{M}(p')$. 
(a) The $s$-box diagram, (b) the $u$-box diagram, and (c) the cat's ears diagram. }
\end{figure}

For simplicity, we assume that the scalar target is made up of two scalar constituents, $Q_1$ and $Q_2$, 
with mass and charge $(m_{Q_1}, e_{Q_1})$ and $(m_{Q_2}, e_{Q_2})$, respectively.
The Mandelstam variables are defined as $s = (p+q)^2$ and $u = (p-q')^2$.
The loop contribution to the scattering amplitude $\mathscr{M}^{\mu}_{\rm tot}$ is given by%
\footnote{From now on, we drop the superscript $(1+1)$ in $\mathscr{M}^{\mu (1+1)}_{\rm tot}$.} 
\begin{equation}
\label{eq14c}
\mathscr{M}^{\mu}_{\rm loop}= \mathscr{M}^{\mu}_{s}  + \mathscr{M}^{\mu}_{u} + \mathscr{M}^{\mu}_{c},
\end{equation}
where $\mathscr{M}^{\mu}_{s}$ and $\mathscr{M}^{\mu}_{u}$ are the $s$- and $u$-channel amplitudes as shown 
in Figs.~\ref{fig1}(a) and~(b), respectively. 
The diagram shown in Fig.~\ref{fig1}(c) is the diagram of ``cat's ears", which we denote as `$c$-channel' 
amplitude. 
The inclusion of the $c$-channel amplitude is crucial to satisfy the gauge invariance.

In the solvable covariant BS model of $(1+1)$-dimensional scalar field theory, the scattering amplitudes of 
$s$-, $u$-, and $c$-channels in the one-loop approximation are written as
\begin{eqnarray}\label{eq15c}
\mathscr{M}^{\mu}_{s}&=&i e_{Q_1} {\cal N} \int \frac{d^{2}k}{(2\pi)^{2}} 
\frac{(2 k+ q)^{\mu}}{N_{k}N_{k+q}N_{k-\Delta}D_{k-p}} ,
\nonumber\\
\mathscr{M}^{\mu}_{u}&=&i e_{Q_1} {\cal N}\int\frac{d^{2}k}{(2\pi)^{2}}
\frac{(2 k- 2q' +q)^{\mu}}{N_{k}N_{k-q'}N_{k-\Delta}D_{k-p}} ,
\nonumber\\
\mathscr{M}^{\mu}_{c}&=&-i e_{{\bar Q}_2} {\cal N}\int\frac{d^{2}k}{(2\pi)^{2}}
\frac{(2 k - 2p -q)^{\mu}}{N_{k}N_{k-q'}D_{k-p-q}D_{k-p}} ,
\end{eqnarray}
where the denominators are coming from the intermediate scalar propagators shown in Fig.~\ref{fig1}.
Here, $N_{p_1} = p^2_1-m_{Q_1}^{2}+i\epsilon$ and $D_{p_2} = p^2_2 -m_{Q_2}^{2}+i\epsilon$.
The normalization constant ${\cal N}$ includes the coupling constants involved in this reaction. 
The electric charges satisfy the charge conservation, $e_{Q_1} + e_{\bar{Q}_2}=e_{\cal M}$, 
where $e_{\cal M}$ is the charge of the scalar target and $e_{{\bar Q}_j} = - e_{Q_j}$.

While one may perform the manifestly covariant calculations of Eq.~(\ref{eq15c}) using the Feynman 
parametrization, it has technical difficulties in analyzing the pole structures associated with the  
multi-dimensional integral of Feynman parameters. 
On the other hand, the LF calculations in $(1+1)$ dimensions avoid such difficulties since it involves only 
the one-dimensional integral, as we shall show below.

\begin{figure*}[t!]\centering
\includegraphics[width=1 \textwidth]{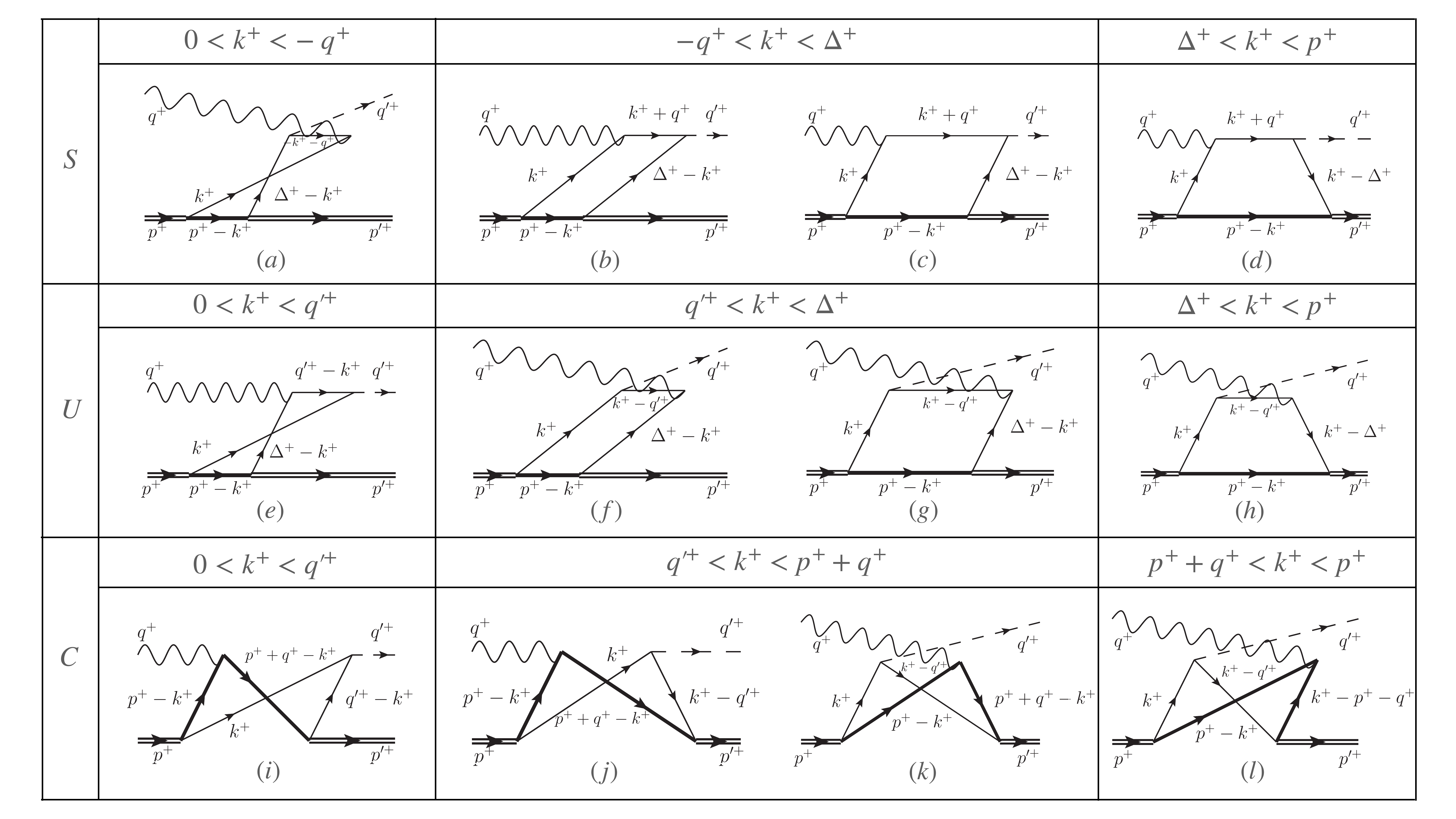} 
\caption{\label{fig2}%
The LF time-ordered diagrams for the scattering amplitudes 
($\mathscr{M}^{\mu}_{s}, \mathscr{M}^{\mu}_{u},\mathscr{M}^{\mu}_{c}$)
corresponding to the $(s, u, c)$-channels.
}
\end{figure*}

In terms of the LF variables, the $s$-channel amplitude $\mathscr{M}^{\mu}_{s}$ in Eq.~(\ref{eq15c}) can be 
rewritten as 
\begin{eqnarray}
\label{eq16c}
\mathscr{M}^{\mu}_{s} &=& \frac{i e_{Q_1} {\cal N}}{2 (2\pi)^2}
\int \frac{dk^{+}dk^{-}}{C_s} \nonumber\\
&& \mbox{} \times
\frac{2 k^{\mu}+q^{\mu}}{(k^{-}-k^{-}_{i}) (k^{-}-k^{-}_{f}) (k^{-}-k^{-}_{t}) (k^{-}-k^{-}_{b})},
\end{eqnarray}
where $C_{s}=k^{+} (k^{+}+q^{+}) (k^{+}-\Delta^{+}) (k^{+}-p^{+})$ and
\begin{eqnarray}
\label{eq17c}
k^{-}_{i} &=& \frac{m_{Q_1}^2}{k^{+}}-i \frac{\epsilon}{k^{+}}, \nonumber\\
k^{-}_{f} &=& \Delta^{-}+\frac{m_{Q_1}^2}{k^{+}-\Delta^{+}}-i \frac{\epsilon}{k^{+}-\Delta^{+}}, \nonumber\\
k^{-}_{t} &=& -q^{-}+\frac{m_{Q_1}^2}{k^{+}+q^{+}}-i \frac{\epsilon}{k^{+}+q^{+}}, \nonumber\\
k^{-}_{b} &=& p^{-}+\frac{m_{Q_2}^2}{k^{+}-p^{+}}-i \frac{\epsilon}{k^{+}-p^{+}}.
\end{eqnarray}
Similar expressions can be obtained for $\mathscr{M}^{\mu}_{u}$  and $\mathscr{M}^{\mu}_{c}$. 
The obtained twelve LF time-ordered diagrams for the scattering amplitudes 
($\mathscr{M}^{\mu}_{s}, \mathscr{M}^{\mu}_{u},\mathscr{M}^{\mu}_{c}$)
corresponding to $(s, u, c)$-channels are depicted in Fig.~\ref{fig2}.

For the $s$-channel amplitude $\mathscr{M}^{\mu}_{s}$ in Eq.~(\ref{eq16c}), the Cauchy integration over $k^-$
gives the three LF time ($x^+$)-ordered contributions to the residue calculations, i.e., those coming from 
regions $S1$ ($\Delta^+ < k^+ < p^+$),  $S2$ ($-q^+ < k^+ <\Delta^+$), and  $S3$ ($0< k^+ < -q^+$), 
respectively. 
There are several comments in order.
We note that $q^+$ in Eq.~(\ref{eq6c}) is chosen to be $q^+<0$ and the region $S2$ is absent for $\mu_s =0$ 
limit as in the case of DVCS. 
The kinematic region $S1$ corresponds to the so-called ``Dokshitzer-Gribov-Lipatov-Altareili-Parisi (DGLAP)''
region~\cite{GL72a,Dokshitzer77,AP77}, and the other two regions, $S2$ and $S3$, correspond to the so-called 
``Efremov-Radyushkin-Brodsky-Lepage (ERBL)'' region~\cite{ER80,LB79,LB80}.
The DGLAP and ERBL regions correspond to the valence contribution representing the particle-number-conserving
process and the nonvalence one representing the particle-number-changing process, respectively.

In the DGLAP region of $S1$, where $\Delta^+ < k^+ < p^+$, the residue is at the pole of $k^-= k^-_b$, which 
is placed in the upper half of the complex $k^-$ plane.  
Therefore, the Cauchy integration of $\mathscr{M}^{\mu}_{s}$ in Eq.~(\ref{eq16c}) over $k^-$ in this region 
leads to
\begin{equation}\label{eq18c}
\mathscr{M}^{\mu}_{s,\rm hand}=-\frac{e_{Q_1} {\cal N}}{4\pi} \int_{\Delta^{+}}^{p^{+}} dk^{+}
\frac{2 k^{\mu}_{b}+q^{\mu}}{C_{s} (\Delta k^-_{bi})(\Delta k^-_{bf}) (\Delta k^-_{bt})},
\end{equation}
where $k^\mu_{j}=(k^+, k^-_{j})$ and $\Delta k^-_{jk}= k^-_j - k^-_k$.
This amplitude corresponds to the ``handbag'' diagram shown in Fig.~\ref{fig2}(d) in the $s$-channel.

In the ERBL region of $S2$, where $-q^+ < k^+ <\Delta^+$, while two poles ($k^{-}_{i}, k^{-}_{t}$) are placed
in the lower half of the complex $k^-$ plane, the other two poles ($k^{-}_{f}, k^{-}_{b}$) lie on the upper 
half of the complex $k^-$ plane.  
Taking the two poles $k^-=(k^{-}_{f}, k^{-}_{b})$ and using some mathematical manipulations for the 
denominators, e.g., $1/ (\Delta k^-_{fi} \Delta k^-_{fb}) = -(1/\Delta k^-_{bi}) (1/\Delta k^-_{fi} 
+ 1/\Delta k^-_{bf})$, we obtain two different types of LF time-ordered amplitudes in the $S2$ region as 
\begin{eqnarray}
\label{eq19c}
\mathscr{M}^{\mu}_{s, \rm stret} &=& \frac{e_{Q_1} {\cal N}}{4\pi} \int^{\Delta^{+}}_{-q^{+}} dk^{+}
\frac{2 k^{\mu}_{i}+q^{\mu}}{C_{s} (\Delta k^-_{bi})(\Delta k^-_{fi}) (\Delta k^-_{ft})},
\nonumber\\
\mathscr{M}^{\mu}_{s, \rm open} &=& \frac{e_{Q_1} {\cal N}}{4\pi} \int^{\Delta^{+}}_{-q^{+}} dk^{+}
\frac{2 k^{\mu}_{b} + q^{\mu}}{C_{s} (\Delta k^-_{bi})(\Delta k^-_{bt}) (\Delta k^-_{ft})},
\end{eqnarray}
where $\mathscr{M}^{\mu}_{s,\rm stret}$ and $\mathscr{M}^{\mu}_{s,\rm open}$ correspond to the so-called 
``stretched box'' and ``open diamond'' diagrams shown in Figs.~\ref{fig2}(b) and~ \ref{fig2}(c) in the 
$s$-channel, respectively.

In the other ERBL region of $S3$, where $0<k^{+}<-q^{+}$, the residue is at the pole of $k^-= k^-_i$, which 
is placed in the lower half of the complex $k^-$ plane.  
The Cauchy integration of $\mathscr{M}^{\mu}_{s}$ in Eq.~(\ref{eq16c}) over $k^-$ in the $S3$ region leads to
\begin{equation}\label{eq20c}
\mathscr{M}^{\mu}_{s,\rm twist}=\frac{e_{Q_1} {\cal N}}{4\pi} \int_{0}^{-q^{+}} dk^{+}
\frac{2 k^{\mu}_{i}+q^{\mu}}{C_{s} (\Delta k^-_{ib})(\Delta k^-_{if}) (\Delta k^-_{it})},
\end{equation}
which corresponds to what we call ``twisted stretched box'' diagram as shown in Fig.~\ref{fig2}(a) in 
the $s$-channel.

Similarly, we can obtain the LF time-ordered amplitudes for $\mathscr{M}^{\mu}_{u}$ of 
Figs.~\ref{fig2}(e)-(h) in the $u$-channel and $\mathscr{M}^{\mu}_{c}$ of Figs.~\ref{fig2}(i)-(l) in 
the $c$-channel.
Their explicit expressions read
\begin{eqnarray}\label{eq21c}
\mathscr{M}^{\mu}_{u,\rm twist} &=& \frac{e_{Q_1}{\cal N}}{4\pi} \int_{0}^{q'^{+}} dk^{+}
\frac{2 k^{\mu}_{i}-2 q'^{\mu}+q^{\mu}}{C_{u}(\Delta k^-_{if})(\Delta k^-_{iu})(\Delta k^-_{ib})},
\nonumber\\
\mathscr{M}^{\mu}_{u,\rm stret} &=& -\frac{e_{Q_1}{\cal N}}{4\pi} \int_{q'^{+}}^{\Delta^{+}} dk^{+}
\frac{2 k^{\mu}_{i}-2 q'^{\mu}+q^{\mu}}{C_{u}(\Delta k^-_{ib})(\Delta k^-_{fi})(\Delta k^-_{fu})},
\nonumber\\
\mathscr{M}^{\mu}_{u,\rm open} &=& -\frac{e_{Q_1}{\cal N}}{4\pi}  \int_{q'^{+}}^{\Delta^{+}} dk^{+}
\frac{-2 k^{\mu}_{b}+2 q'^{\mu}-q^{\mu}}{C_{u}(\Delta k^-_{ub})(\Delta k^-_{fu})(\Delta k^-_{ib})},
\nonumber\\
\mathscr{M}^{\mu}_{u,\rm hand} &=& -\frac{e_{Q_1}{\cal N}}{4\pi} \int_{\Delta^{+}}^{p^{+}} dk^{+}
\frac{2 k^{\mu}_{b}-2 q'^{\mu}+q^{\mu}}{C_{u} (\Delta k^-_{bf})(\Delta k^-_{bu})(\Delta k^-_{bi})}, 
\nonumber\\
\end{eqnarray}
and 
\begin{eqnarray}\label{eq22c}
\mathscr{M}^{\mu}_{c(i)} &=& -\frac{e_{{\bar Q}_2}{\cal N}}{4\pi} \int_{0}^{q'^{+}} dk^{+}
\frac{2 k^{\mu}_{i}-2 p^{\mu}-q^{\mu}}{C_{c} (\Delta k^-_{ib})(\Delta k^-_{ic})(\Delta k^-_{iu}) },
\nonumber\\
\mathscr{M}^{\mu}_{c(j)} &=& \frac{e_{{\bar Q}_2}{\cal N}}{4\pi} \int_{q'^{+}}^{p^{+}+q^{+}} dk^{+}
\frac{-2 k^{\mu}_{i}+2 p^{\mu}+q^{\mu}}{C_{c} (\Delta k^-_{bi})(\Delta k^-_{ci})(\Delta k^-_{cu})},
\nonumber\\
\mathscr{M}^{\mu}_{c(k)} &=& \frac{e_{{\bar Q}_2}{\cal N}}{4\pi} \int_{q'^{+}}^{p^{+}+q^{+}} dk^{+}
\frac{-2 k^{\mu}_{b}+2 p^{\mu}+q^{\mu}}{C_{c}(\Delta k^-_{bi})(\Delta k^-_{bu})(\Delta k^-_{cu})},
\nonumber\\
\mathscr{M}^{\mu}_{c(l)} &=& \frac{e_{{\bar Q}_2}{\cal N}}{4\pi} \int_{p^{+}+q^{+}}^{p^{+}} dk^{+}
\frac{2 k^{\mu}_{b}-2 p^{\mu}-q^{\mu}}{C_{c}(\Delta k^-_{bc})(\Delta k^-_{bu})(\Delta k^-_{bi})},
\end{eqnarray}
where $C_{u}=k^{+} (k^{+}-q'^{+}) (k^{+}-\Delta^{+}) (k^{+}-p^{+})$, $C_{c}=k^{+} (k^{+}-q'^{+}) 
(k^{+}-p^{+}-q^{+}) (k^{+}-p^{+})$, and
\begin{eqnarray}\label{eq23c}
&&k^{-}_{u}=q'^{-}+\frac{m^2_{Q_1}}{k^{+}-q'^{+}}-i \frac{\epsilon}{k^{+}-q'^{+}},\nonumber\\
&&k^{-}_{c}=p^{-}+q^{-}+\frac{m^2_{Q_2}}{k^{+}-p^{+}-q^{+}}-i \frac{\epsilon}{k^{+}-p^{+}-q^{+}}.
\end{eqnarray}

\subsection{Amplitudes from Effective Tree Diagrams}
\label{sec:3-B}

\begin{figure}[t!]\centering
\includegraphics[width=0.5\textwidth]{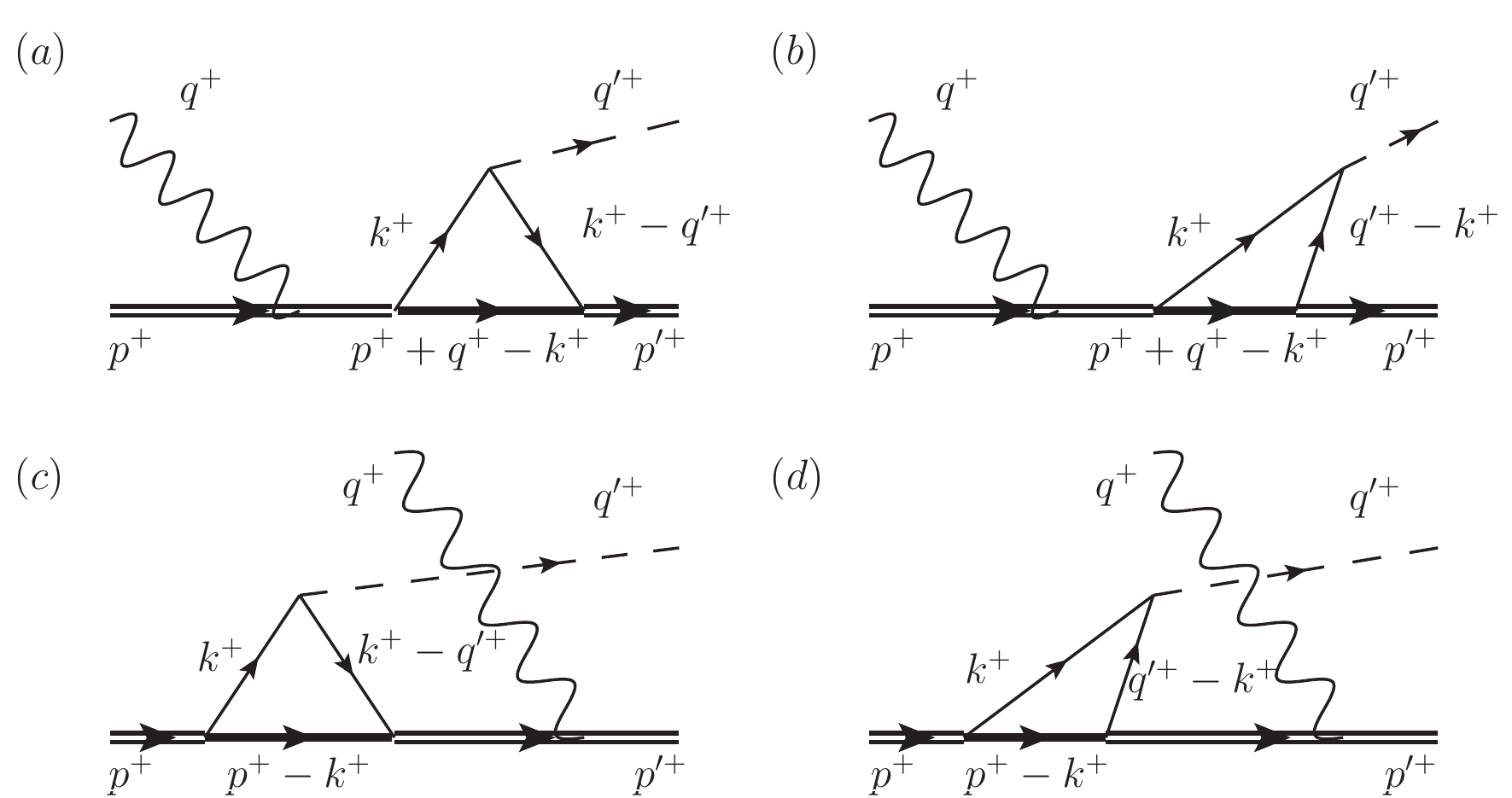}
\caption{\label{fig3}%
 The LF time-ordered effective tree diagrams in $s$- and $u$-channels for a charged target case.
(a,b) The valence and nonvalence contributions to $\mathscr{M}^{\mu}_{s, \rm ET}$, respectively,
and
(c,d) the valence and nonvalence contributions to $\mathscr{M}^{\mu}_{u, \rm ET}$, respectively.
}
\end{figure}

For the neural target, where $e_{\cal M}=e_{Q_1}+e_{{\bar Q}_2}=0$, the gauge invariance condition
$q \cdot \mathscr{M}^{\mu}_{\rm tot}=0$ is guaranteed when 
$\mathscr{M}^{\mu}_{\rm tot}= \mathscr{M}^{\mu}_{\rm loop}$. 
However, for the case of a charged target such as the ``helium" nucleus, additional diagrams 
called ``effective tree'' diagrams, where the photon line is attached to the charged target, are required to 
ensure the gauge invariance.
The effective tree contribution to the scattering amplitude is decomposed as
\begin{equation}
\label{eq24c}
\mathscr{M}^{\mu}_{\rm ET}=\mathscr{M}^{\mu}_{s,\rm ET} + \mathscr{M}^{\mu}_{u, \rm ET},
\end{equation} 
where the corresponding LF time-ordered diagrams are presented in Fig.~\ref{fig3}. 
The covariant scattering amplitudes $\mathscr{M}^{\mu}_{s,\rm ET}$ and $\mathscr{M}^{\mu}_{u, \rm ET}$ are 
obtained as
\begin{eqnarray}
\label{eq25c}
\mathscr{M}^{\mu}_{s, \rm ET} &=& \frac{i e_{\cal M} {\cal N}}{(p+q)^2-{M_{T}}^{2}}
\int \frac{d^{2}k}{(2 \pi)^2}\frac{2 p^{\mu}+q^{\mu}}{N_{k} N_{k-q'} D_{k-p-q}},
\nonumber\\
\mathscr{M}^{\mu}_{u, \rm ET} &=& \frac{i e_{\cal M} {\cal N} }{(p-q')^2-{M_{T}}^{2}}
\int \frac{d^{2}k}{(2 \pi)^2}\frac{2 p^{\mu}+q^{\mu}-2 q'^{\mu}}{N_{k} N_{k-q'} D_{k-p}}.
\nonumber \\
\end{eqnarray}

In the LF calculations, the Cauchy integration over $k^-$ in Eq.~(\ref{eq25c}) gives two LF time-ordered 
contributions to the residue calculations.
For $\mathscr{M}^{\mu}_{s, \rm ET}$, one comes from the valence region $(q'^{+}<k^{+}<p^{+}+q^{+})$ as shown 
in Fig.~\ref{fig3}(a) and the other from the nonvalence region $(0<k^{+}<q'^{+})$ as shown in 
Fig.~\ref{fig3}(b).
In the case of $\mathscr{M}^{\mu}_{u, \rm ET}$, they come from the valence region $(q'^{+}<k^{+}<p^{+})$ as 
shown in Fig.~\ref{fig3}(c) and from the nonvalence region $(0<k^{+}<q'^{+})$ as shown in Fig.~\ref{fig3}(d).
In the valence (nonvalence) region for $\mathscr{M}^{\mu}_{s,\rm ET}$, the residue is at the pole of 
$k^- = k^-_c (k^-_i)$, which is placed in the upper (lower) half of the complex $k^-$ plane. 
Similarly, in the valence (nonvalence) region for $\mathscr{M}^{\mu}_{u, \rm ET}$, the residue is at the pole
of $k^- = k^-_b (k^-_i)$, which is placed in the upper (lower) half of the complex $k^-$ plane.
Thus, the Cauchy integrations of $\mathscr{M}^{\mu}_{s(u), \rm ET}$ over $k^-$ lead to
\begin{eqnarray}
\label{eq26c}
\mathscr{M}^{\mu}_{s,\rm ET} &=& \frac{e_{\cal M} {\cal N}}{ 4\pi [(p+q)^2-{M_{T}}^{2}]}
\nonumber \\ && \mbox{} \times
\left[  \int_{0}^{q'^{+}} dk^{+}\frac{2 p^{\mu}+q^{\mu}}{C'_{s} (\Delta k^-_{iu})(\Delta k^-_{ic})}
\right. \nonumber\\\nonumber\\ && \qquad \left. 
-\int_{q'^{+}}^{p^{+}+q^{+}} dk^{+}\frac{2 p^{\mu}+q^{\mu}}{C'_{s} (\Delta k^-_{ci})(\Delta k^-_{cu})} 
\right],
\nonumber\\
\mathscr{M}^{\mu}_{u, \rm ET} &=& \frac{e_{\cal M} }{ 4\pi[(p-q')^2-{M_{T}}^{2}]}
\nonumber \\ && \mbox{} \times
\left[ \int_{0}^{q'^{+}} dk^{+} \frac{2 p^{\mu}+q^{\mu}-2 q'^{\mu}}{C'_{u} (\Delta k^-_{iu})
(\Delta k^-_{ib})} 
\right. \nonumber\\\nonumber\\ && \qquad \left. 
- \int_{q'^{+}}^{p^{+}} dk^{+} \frac{2 p^{\mu}+q^{\mu}-2 q'^{\mu}}{C'_{u} (\Delta k^-_{bi})(\Delta k^-_{bu})}
\right],
\label{eq:wideeq}
\end{eqnarray}
where~ $C'_{s}=k^{+} (k^{+}-q'^{+}) (k^{+}-p^{+}-q^{+})$ and $C'_{u}=k^{+} (k^{+}-q'^{+}) (k^{+}-p^{+})$.

It should be noted that the full amplitudes are obtained by including the exchanged diagrams 
$(Q_1\leftrightarrow Q_2)$ in Figs.~\ref{fig1}-\ref{fig3}.
Although we do not give the corresponding amplitudes explicitly, their expressions can straightforwardly be obtained 
from the formulae given above with the exchange of $Q_1\leftrightarrow Q_2$.
It should be understood that the contributions from the exchange of 
$Q_1\leftrightarrow Q_2$ are included in our numerical computation of
the full amplitudes even if they are not explicitly mentioned.
Thus, the total scattering amplitudes for the neutral and charged targets may be summarized as 
\begin{eqnarray}
\label{eq27c}
\mathscr{M}^{\mu}_{\rm neutral} &=& \mathscr{M}^{\mu}_{\rm loop}(Q_1,Q_2) +  (1 \leftrightarrow 2),
\nonumber\\
\mathscr{M}^{\mu}_{\rm charged} &=&
\mathscr{M}^{\mu}_{\rm loop}(Q_1, Q_2)+\mathscr{M}^{\mu}_{\rm ET}(Q_1,Q_2) + (1 \leftrightarrow 2),
\nonumber\\
\end{eqnarray}
respectively. 
The CFF $\mathcal{F}_{c}$ for the charged target such as the ``helium" nucleus is then computed by
\begin{equation}
\label{eq28c}
\mathcal{F} ^{\rm VMP}_{c}(Q^2, t)
= \frac{\mathscr{M}^{\mu}_{\rm charged}}{(\Delta\cdot q) q^{\mu}- q^2 \Delta^{\mu}},
\end{equation}
which is valid for each component ($\mu=+,-$) of the current
in $(1+1)$ dimensions. 

\section{Deeply Virtual Meson Production Limit}
\label{sec:4}

In this section, we analyze the amplitude in the DVMP limit, where $Q^2$ is larger than the other scales, 
namely,  $Q^{2} \gg (M^2_T, M^2_S, -t)$.
From $Q^{2} \gg (M^2_S, -t)$, we have $(\mu_s, \tau) \to 0$, which leads to $\zeta = \zeta'$, 
$q^- = q'^- = Q^2/\zeta p^+$, $q^+ = -\Delta^+$, and $q'^+ = 0$ from Eqs.~(\ref{eq6c}) and (\ref{eq7c}).
Furthermore, we also have $\Delta k^-_{bt}=\Delta k^-_{ft}=\Delta k^-_{it} = q^-$ in the energy denominators 
for the scattering amplitudes given by Eqs.~(\ref{eq18c})-(\ref{eq20c}).
However, it should be noted that the condition $Q^2 \gg M^2_T$ is not used here in taking the DVMP limit.

\subsection{Generalized Parton Distribution}\label{sec:4-A}

In the DVMP limit, the time-ordered amplitudes for the $s$-channel with $\mu=+$ given by 
Eqs.~(\ref{eq18c})-(\ref{eq20c}) are now reduced to 
\begin{eqnarray}
\label{eq29c}
\mathscr{M}^{+\rm DVMP}_{s,\rm hand} &=&\frac{e_{Q_1} {{\cal N}}}{4\pi q^-} \int_{\Delta^{+}}^{p^{+}} dk^{+}
\frac{-2 k^+ + \Delta^+}{C_{s} (k^{-}_{b}-k^{-}_{i}) (k^{-}_{b}-k^{-}_{f})},
\nonumber\\
\mathscr{M}^{+\rm DVMP}_{s,\rm twist} &=&\frac{e_{Q_1} {{\cal N}}}{4\pi q^-} \int_{0}^{\Delta^{+}} dk^{+}
\frac{2 k^+ - \Delta^+ }{C_{s} (k^{-}_{i}-k^{-}_{b}) (k^{-}_{i}-k^{-}_{f})},
\end{eqnarray}
at the  leading order in $Q^2$, where we have used $q^+ =-\Delta^+$. 
The kinematic region for both the ``open diamond'' and  ``stretched box'' diagrams given by Eq.~(\ref{eq19c}) vanishes in the limit of 
$q^+ = -\Delta^+$. 
Also, the ``effective tree'' amplitude $\mathscr{M}^{+\rm DVMP}_{s,\rm ET}$ in Eq.~(\ref{eq26c}) does not contribute in the DVMP limit.

Similarly, the reduced amplitudes for the $u$-channel in the DVMP limit are given by
\begin{eqnarray}
\label{eq30c}
\mathscr{M}^{+\rm DVMP}_{u,\rm hand} &=& \frac{e_{Q_1} {{\cal N}}}{4\pi q^-} 
\int_{\Delta^{+}}^{p^{+}} dk^{+}\frac{2 k^+ - \Delta^+}{C_{u} (k^{-}_{b}-k^{-}_{i}) (k^{-}_{b}-k^{-}_{f})},
\nonumber\\
\mathscr{M}^{+\rm DVMP}_{u,\rm stret} &=& \frac{e_{Q_1} {{\cal N}}}{4\pi q^-} 
\int_{0}^{\Delta^{+}} dk^{+}\frac{-2 k^+ + \Delta^+ }{C_{u} (k^{-}_{i}-k^{-}_{b}) (k^{-}_{i}-k^{-}_{f})}
\end{eqnarray}
from Eq.~(\ref{eq21c}).
Neither the ``effective tree'' amplitude $\mathscr{M}^{+\rm DVMP}_{u,\rm ET}$ in Eq.~(\ref{eq26c}) nor the 
amplitudes of the ``cat's ears" in Eq.~(\ref{eq22c}) contribute in the DVMP limit.

\begin{figure}[t]\centering
\includegraphics[width=8cm, height=3cm]{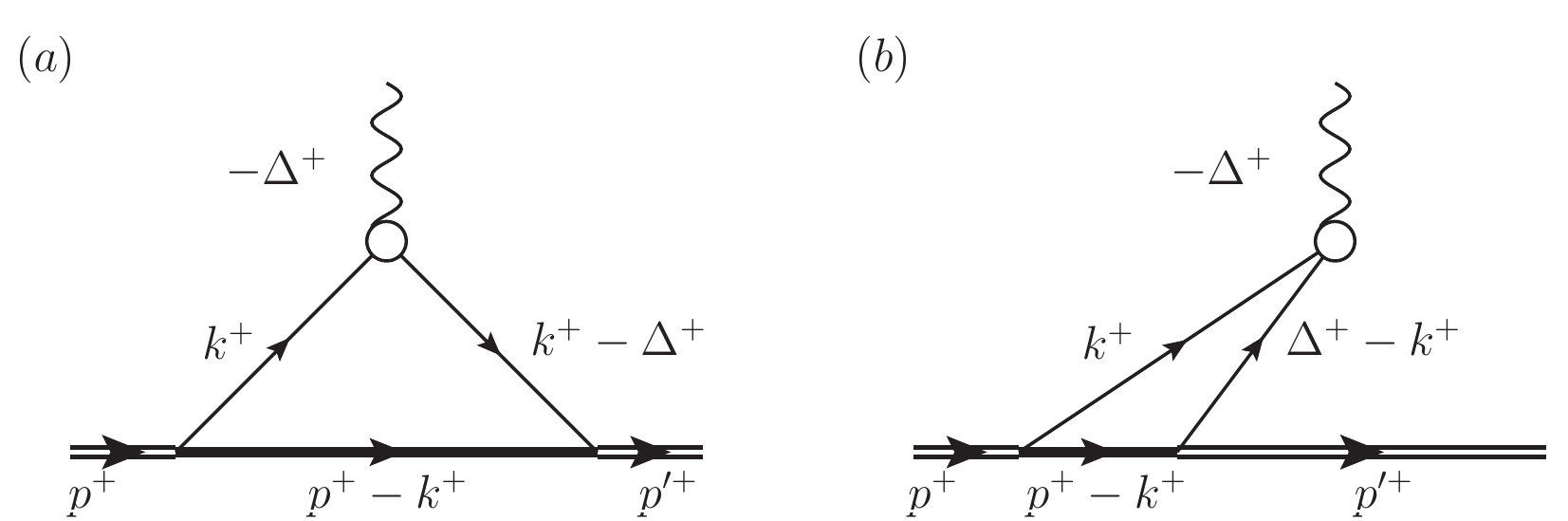}
\caption{\label{fig4}%
Diagrams for GPDs in different kinematic regions for $\zeta>0$ case.
The sum of $\mathscr{M}^{+\rm DVMP}_{s,\rm hand}$ and $\mathscr{M}^{+\rm DVMP}_{u,\rm hand}$
corresponds to the valence diagram (a) defined in the DGLAP ($\zeta\leq x\leq 1$) region and the sum of 
$\mathscr{M}^{+\rm DVMP}_{s, \rm twist}$ and $\mathscr{M}^{+\rm DVMP}_{u,\rm stret}$ corresponds to 
the nonvalence diagram (b) defined in the ERBL ($0\leq x \leq \zeta$) region. 
The small white blob in the figure represents the nonlocality of the constituent--gauge-boson vertex.
}
\end{figure}

Combining both $s$- and $u$-channel amplitudes given by Eqs.~(\ref{eq29c}) and (\ref{eq30c}) and using
the longitudinal momentum fraction $x=k^+/p^+$ ($0\leq x\leq 1$), we obtain the DVMP amplitude in the 
leading order of $Q^2$ as the factorized form of the hard and soft parts given by
\begin{equation}
\label{eq31c}
\mathscr{M}^{+\rm DVMP}_{s+u} = \frac{e_{Q_1}\zeta}{4\pi Q^2} \int_{0}^{1} dx  
\left( \frac{1}{x-\zeta}-\frac{1}{x} \right) H(\zeta, x,t),
\end{equation}
where 
\begin{equation}
\label{eq32c}
H(\zeta, x,t)=
\begin{cases}
H_{\rm ERBL}(\zeta, x,t), & \text{for }~ 0\leq x\leq \zeta, \\
H_{\rm DGLAP}(\zeta, x,t), & \text{for }~ \zeta\leq x\leq 1
\end{cases}
\end{equation}
is identified as the GPD~\cite{Ji96a,Ji96b,Radyushkin96a,Radyushkin97b,MRGDH94,GPV01,Diehl03,BR05}.
The GPD function $H(\zeta, x,t)$ is naturally represented by the sum of the LF nonvalence contribution
to the ERBL ($0\leq x\leq \zeta$) region and the valence contribution to the DGLAP ($\zeta\leq x\leq 1$) 
region as shown in Fig.~\ref{fig4}.

Respectively, $H_{\rm ERBL}(\zeta, x,t)$ and $H_{\rm DGLAP}(\zeta, x,t)$ are given by
\begin{eqnarray}
\label{eq33c}
H_{\rm ERBL} &=& \frac{{\cal N}}{x (\zeta-x) (1-x)}
\frac{(2 x-\zeta)\zeta}{\left( M_{T}^{2}-M_{0}^{2} \right) \left(t + 
\frac{\zeta^{2} m_{Q_{1}}^{2}}{x \left( x-\zeta \right)} \right)},
\nonumber\\
H_{\rm DGLAP} &=& \frac{{\cal N}}{x (x-\zeta)(1-x)}
\frac{(2 x-\zeta) (1-\zeta)}{ \left( M_{T}^{2}-M_{0}^{2} \right) \left( M_{T}^{2}-{M}_{0}^{\prime 2} 
\right)},
\end{eqnarray}
where 
\begin{eqnarray}
\label{eq34c}
M_{0}^{2} = \frac{m_{Q_{1}}^{2}}{x}+\frac{m_{Q_{2}}^{2}}{1-x},
\qquad
{M}_{0}^{\prime 2} = \frac{m_{Q_{1}}^{2}} {x^{\prime}}+\frac{m_{Q_{2}}^{2}}{1-x^{\prime}}
\end{eqnarray}
with $x' = (x-\zeta)/(1-\zeta)$.
It can be checked that $H_{\rm ERBL}(\zeta,x,t)$ and $H_{\rm DGLAP}(\zeta,x,t)$ obtained in this model are 
continuous and finite at the boundary $x=\zeta$, namely, 
$H_{\rm ERBL}(\zeta,\zeta,t)=H_{\rm DGLAP}(\zeta,\zeta,t)=H(\zeta,\zeta, t)$,
which is written explicitly as
\begin{equation}
\label{eq35c}
H(\zeta,\zeta, t)
=\frac{ {\cal N} \zeta}{(1- \zeta) m_{Q_1}^4  + \zeta  m_{Q_1}^2 \left[ m_{Q_2}^2+(\zeta -1) M_T^2\right]}.
\end{equation}
It is related with the imaginary part of the DVMP amplitude $\mathscr{M}^{+\rm DVMP}_{s+u}$ in 
Eq.~(\ref{eq31c}).
As we have mentioned in Sec.~\ref{sec:2}, $\zeta$ and $t$ are not independent of each other in ($1+1$) 
dimensions unlike the $(3+1)$ dimensional case and the explicit expression in Eq.(\ref{eq35c}) is given by a function of $\zeta$ only.

In the DVMP limit and at the leading order of $Q^2$, the contributions from ``effective tree'' amplitudes
are suppressed and only the $s$- and $u$-channel loop amplitudes contribute. 
Effectively, the DVMP results given by $\mathscr{M}^{+\rm DVMP}_{s+u}$ are independent of the electric 
charge of the target, whether it is charged or neutral.
Taking into account the corresponding prefactor in Eq.~(\ref{eq13c}) relating the scattering amplitude to the CFF given by
\begin{equation}
\label{eq36b}
(\Delta\cdot q) q^{+}- q^2 \Delta^{+} =\frac{1}{2}Q^2\zeta p^+ \left[ 1 + \frac{t}{Q^2} +\cdots \right],
\end{equation}
we obtain the CFF in the DVMP limit at the leading order of $Q^2$ denoted by $\mathcal{F}^{\rm DVMP}$ as
\begin{equation}
\label{eq36c}
\mathcal{F}^{\rm DVMP} (Q^2, t)
= \frac{\mathscr{M}^{+\rm DVMP}_{s+u}}{\frac{1}{2}Q^2\zeta p^+}.
\end{equation}
In view of the QCD collinear factorization theorem at the leading twist for the DVMP process~\cite{Collins}, 
the comparison of $\mathcal{F}^{\rm DVMP} (Q^2, t)$ of Eq.~(\ref{eq36c}) with the exact results 
$\mathcal{F} _{c}(Q^2, t)$ of Eq.~(\ref{eq28c}) would be very interesting as it allows us to explore 
the valid kinematic region for the GPD formulation based on the handbag dominance in the leading order of $Q^2$.
We compare the numerical results of $\mathcal{F}_{c}$ and the leading twist $\mathcal{F}^{\rm DVMP} (Q^2, t)$ 
in Sec.~\ref{sec:5}.

\subsection{Parton Distribution Functions}
\label{sec:4-B}

In the forward limit $(\zeta, t)\to 0$, we have $H(0, x,0)=H_{\rm DGLAP}(0,x,0)\equiv q_v(x)$, where
\begin{equation}
\label{eq38c}
q_v(x)=\frac{2 \mathcal{N}}{x (1-x) (M^2_T - M^2_0)^{2}}.
\end{equation}
In this limit, $H_{\rm ERBL}(0,x,0)=0$ and $q_v(x)$ corresponds to the ordinary PDF representing 
the probability to find the constituent inside the hadron as a function of the momentum fraction $x$ carried 
by the constituent in the valence sector. 
The corresponding LF wave function $\psi(x)$ of the target hadron in the momentum space  may be written as
\begin{eqnarray}\label{eq39c}
\psi(x)=\sqrt{q_v (x)}, 
\end{eqnarray}
which satisfies $\int^1_0 dx\, \abs{\psi(x)}^2  = \frac{1}{2}\int_{-1}^{1} dy \, q_v(y)=1$. 
We then obtain the $n$-th moment of $q_v(x)$ defined by~\cite{CZ84}
\begin{equation}
\label{eq40c}
\Braket{y_n} = \int_{-1}^{1} dy\, y^{n} q_v(y),
\end{equation}
where $y=2 x-1$.

Introducing the longitudinally boost-invariant dimensionless LF spatial variable $\tilde{z}=p^{+}x^{-}$,
which is canonically conjugate to $x$~\cite{MB19,BD07a,HLFHS18},
the LF wave function $\psi(\tilde{z})$ in the LF coordinate space $\tilde{z}$ evaluated at $x^+=0$ can be 
defined by
\begin{equation}
\label{eq41c}
\psi(\tilde{z}) = \frac{1}{\sqrt{2\pi}} \int dx \, \psi (x) e^{i k\cdot x} = \frac{1}{\sqrt{2\pi}} 
\int_{0}^{1} dx \, \psi (x) e^{i\tilde{z}x/2}
\end{equation}
as the Fourier transform of $\psi(x)$ in $(1+1)$ dimensions.
Then, the longitudinal probability density $\varrho (\tilde{z})$ in the LF coordinate space $\tilde{z}$ is 
given by
\begin{equation}\label{eq42c}
\varrho (\tilde{z})=\vert\psi(\tilde{z})\vert^{2},
\end{equation}
which satisfies $\int_{0}^{\infty} \varrho (\tilde{z}) \, d\tilde{z}=1$.  
Detailed discussions on the three dimensional version of Eq.~(\ref{eq41c}), $\varrho (\tilde{z}, \mathbf{b})$,
which includes the transverse distance $\mathbf{b}$ of the struck constituent from the transverse center of momentum were provided in Refs.~\cite{MB19,BD07a,HLFHS18}.

\subsection{Moments of GPD}
\label{sec:4-C}

In general, the $n$-th moment of the GPD is defined by
\begin{equation}
\label{eq43c}
F_n (\zeta, t)=\int^1_0 \frac{dx}{1-\zeta/2} \, x^{n-1} H(\zeta, x, t).
\end{equation} 
It is well known that the polynomiality conditions~\cite{JMS97,Radyushkin98} for the $n$-th moment of the GPD 
require that the highest power of $\zeta$ in the polynomial expression of $F_n(\zeta, t)$ should not be larger than $n$. 
These polynomiality conditions are fundamental properties of the GPD, which follow from the Lorentz invariance.

The first moment of $H(\zeta, x,t)$ is related to the EM form factor $F_{\mathcal{M}} (t)$ of the target ${\mathcal{M}}$ 
by the following sum rule~\cite{Ji96a,Ji96b,Radyushkin96a,Radyushkin97b}:
\begin{equation}
\label{eq44c}
F_{\mathcal{M}} (t) = \int^1_0 \frac{dx}{1-\zeta/2}  H(\zeta, x, t).
\end{equation}
In the $(3+1)$ dimensional analysis,  the full result of the EM form factor ($n=1$) should be independent of 
$\zeta$ so that $F_1 (\zeta, t) = F_{\mathcal{M}}(t)$ since the two variables $\zeta$ and $t$ are independent of 
each other.
However, in $(1+1)$ dimensions, the moment $F_n(\zeta, t)$ should be a function of a single variable, 
$F_n(\zeta)$ or $F_n(t)$, since the two variables are related to each other.
For example, $\zeta=0$ and 1 correspond to $-t=0$ and $\infty$, respectively.
In other words, the interval of $\zeta=[0, 1]$ covers the entire range of the momentum transfer squared $-t = [0,\infty]$ in the $n$-th moment of the GPD in $(1+1)$ dimensions.
Furthermore, all the moments vanish at $\zeta=1$, i.e., $F_n (\zeta=1)=0$, which hinders to check the 
polynomiality conditions. 
To circumvent this problem in checking the polynomiality conditions due to $F_n (\zeta=1)=0$, we redefine 
the moments as
\begin{eqnarray}\label{eq45c}
{\bar F}_{n}(\zeta)=F_{n}(\zeta)/F_{1}(\zeta),
\end{eqnarray}
so that  $\bar{F}_{1}(\zeta)$ is independent of $\zeta$. 
In Sec.~\ref{sec:5}, we will discuss how our model calculations for $\bar{F}_{n}(\zeta)$ satisfy the 
polynomiality conditions.

The normalization factor ${\cal N}$ is fixed by the condition $F_{\mathcal{M}}(0)=1$ and given by
\begin{eqnarray}\label{eq46c}
{\cal N} =\frac{m_{Q_1}^2~ m_{Q_2}^2~ (1-\omega^2)^2}{1-\omega^2 +\omega \sqrt{1-\omega^{2}}~ T_{\omega}},
\end{eqnarray}
where
\begin{equation}\label{eq47c}
\omega = \frac{M_{T}^{2}-m_{Q_2}^{2}-m_{Q_1}^{2}}{2 m_{Q_1} m_{Q_2}},
\end{equation}
and
\begin{equation}
\label{eq48c}
T_{\omega} = {\rm tan}^{-1}\left(\frac{m_{Q_2}+m_{Q_1} \omega}{m_{Q_1} \sqrt{1-\omega^2}}\right)
+{\rm tan}^{-1}\left(\frac{m_{Q_1} +m_{Q_2} \omega}{m_{Q_2} \sqrt{1-\omega^2}}\right).
\end{equation} 
We note that the EM form factor obtained by using Eqs.~(\ref{eq33c}) and~(\ref{eq44c}) is identical to 
the form factor obtained in our previous publication~\cite{CCJO21} within the same solvable $\phi^3$ scalar 
field model in $(1+1)$ dimensions~\cite{SM88,MS89,GS90}.

\section{Numerical results} 
\label{sec:5}

For the numerical computation, we simulate the electroproduction of a scalar meson off the scalar target \nuclide[4]{He} with 
the electric charge $e_{\cal M}=+2e$ using the ($1+1$) dimensional scalar field theory discussed in the present work.
For our numerical calculations and analyses, we thus take the target and scalar meson masses as  
$M_{T}=3.7$~GeV and $M_{S}=0.98$~GeV. 
In this case, the threshold momentum transfer at $Q^2=0$ is given by 
$t_{\rm th}(Q^{2}=0) = - M_T M^2_S / (M_T+ M_S) \simeq -0.76~\mbox{GeV}^{2}$.
For the constituent mass, we use the equal mass for both constituents, $m_{Q_1} = m_{Q_2} = 2~\mbox{GeV}$, 
so that the \nuclide[4]{He} target is a weakly bound state, as 
$M_T < m_{Q_1} + m_{Q_2}$ but  $M_{T}^{2} > m_{Q_{1}}^{2} + m_{Q_{2}}^{2}$.%
\footnote{Therefore, we mimic the reaction of $f_0(980)$ production off the \nuclide[4]{He} targets in 
$(1+1)$ dimensions assuming that the \nuclide[4]{He} nucleus is a bound state of two (scalar) deuterons.}
However, we will discuss the cases with some variations of constituent masses as needed for comments.
For the consistency of our numerical analysis, we use the same normalization constant ${\cal N}$ given by 
Eq.~(\ref{eq46c}) for all physical observables 
such as the CFF $\mathcal{F}(Q^2,t)$, the GPD $H(\zeta, x, t)$, and the EM form factor $F_{\cal M}(t)$ 
throughout the present work.

\subsection{CFF in VMP}
\label{sec:5-A}

\begin{figure}[t!]\centering
\includegraphics[width=0.8\columnwidth]{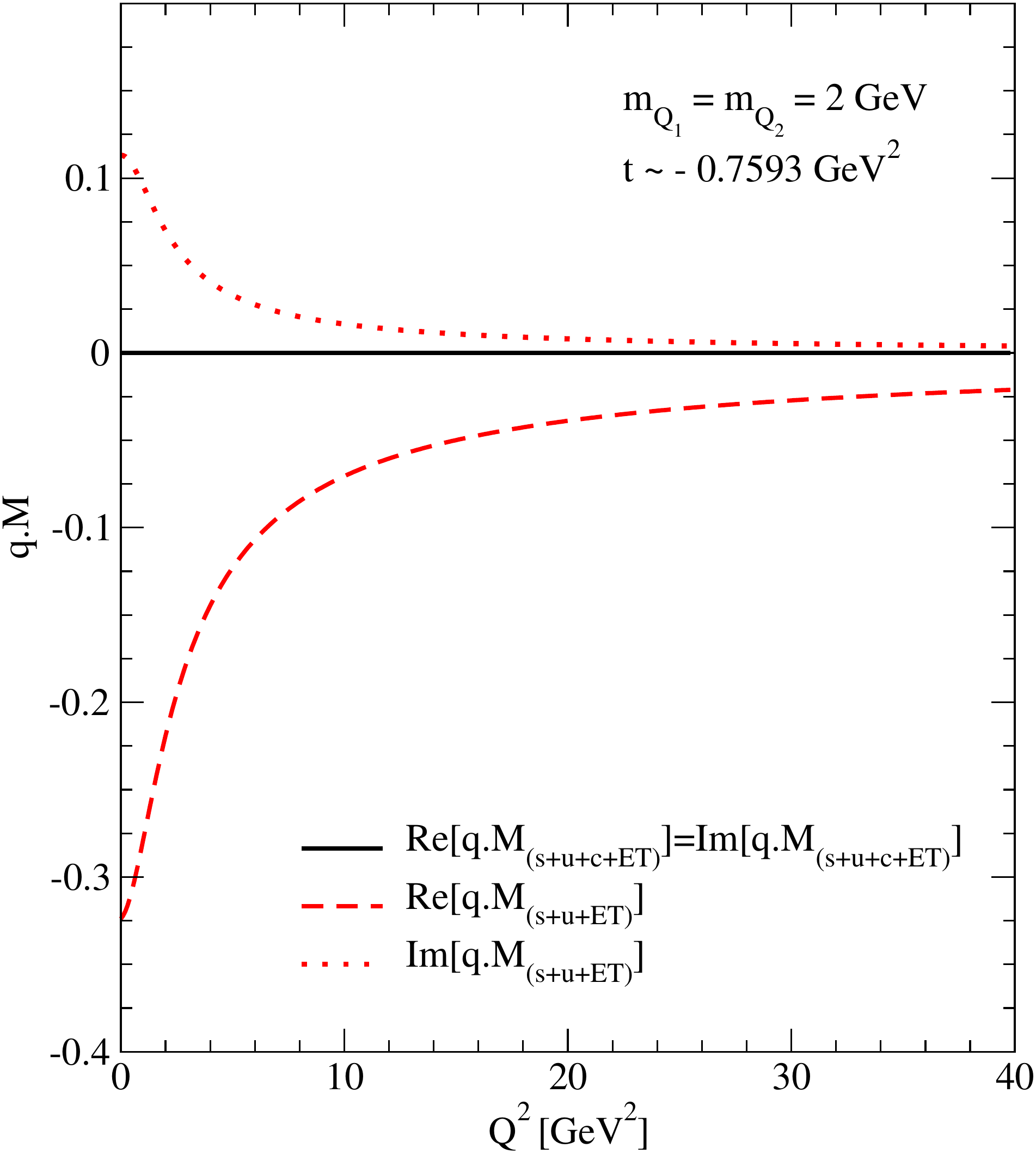}
\caption{\label{fig5}%
Role of the $c$-channel contributions in the gauge invariance condition of $\mbox{Re} \left[ q \cdot \mathscr{M} \right]$ 
and $\mbox{Im} \left[ q \cdot \mathscr{M} \right]$ for a charged target. 
The solid line is for the full amplitude while the dashed and dotted lines are for the amplitudes without 
the $c$-channel contributions.
}
\end{figure}

As we have discussed before, the total scattering amplitude for a charged target $\mathscr{M}^\mu=\mathscr{M}^\mu_{\rm charged}$ 
given by Eq.~(\ref{eq27c}), i.e., the full results without any approximation, should satisfy the condition that 
$q \cdot \mathscr{M}=0$. 
Furthermore, since $\mathscr{M}^\mu$ is in general a complex-valued function, even for the spacelike region 
$Q^2>0$, the real and imaginary part of $\mathscr{M}^\mu$ can be shown to satisfy the gauge invariance condition
separately, i.e., ${\rm Re}[q\cdot\mathscr{M}]= 0$ and ${\rm Im}[q\cdot\mathscr{M}]= 0$.
We first check numerically whether these conditions are met by our amplitudes.
The solid line of Fig.~\ref{fig5} shows that the gauge invariance is observed by the exact amplitude 
$\mathscr{M}^\mu_{\rm charged} =\mathscr{M}^\mu_{u+s+c+\rm ET}$ for the range of $0\leq Q^2\leq 40$ GeV$^2$ at 
$t=t_{\rm th}(Q^2=0)$. 
To estimate the $c$-channel contribution, we turn off the $c$-channel contributions, and the results for the 
real and imaginary parts of $q\cdot\mathscr{M}_{u+s+\rm ET}$ are plotted by the dashed and dotted lines, 
respectively, in Fig.~\ref{fig5}.
This evidently shows that the omission of the ``cat's ears" diagrams violates the Ward identity. 
The violation is more serious at the smaller $Q^{2}$ region, although the degree of deviation weakens at the 
larger $Q^2$ region as anticipated.

\begin{figure*}[t]\raggedright
\includegraphics[width=5.6cm, height=5.6cm]{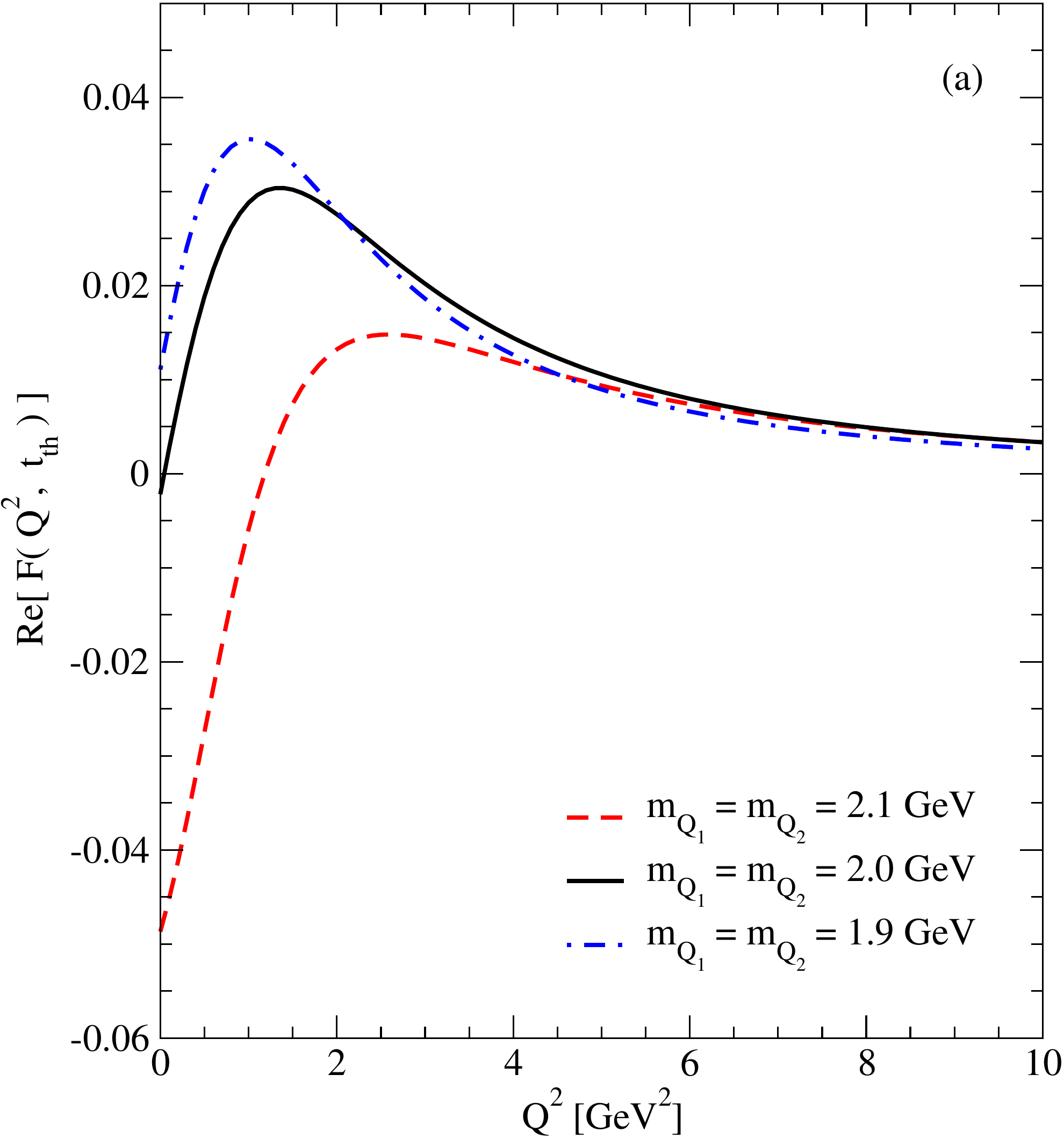}
\includegraphics[width=5.6cm, height=5.6cm]{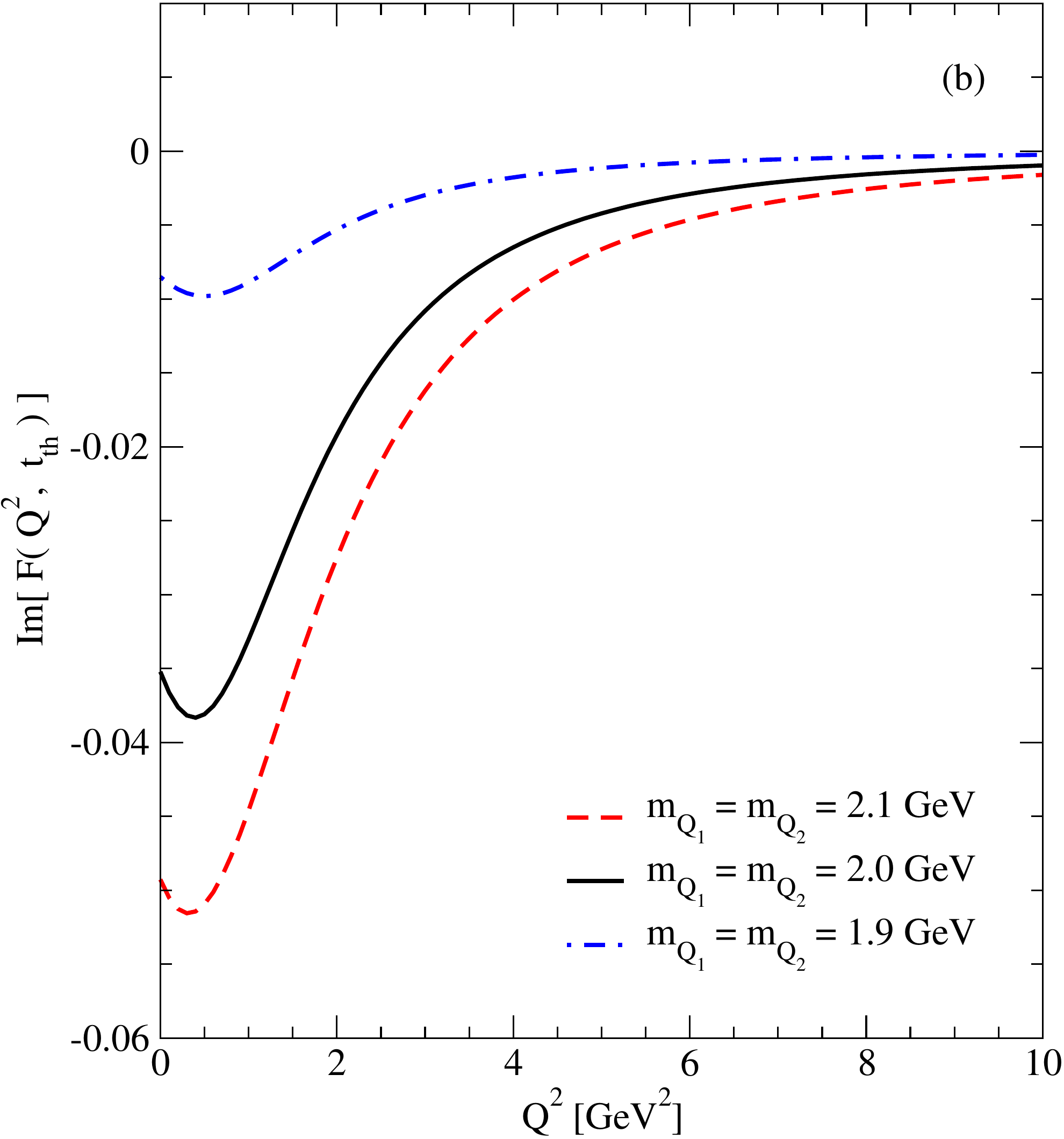}
\includegraphics[width=5.6cm, height=5.6cm]{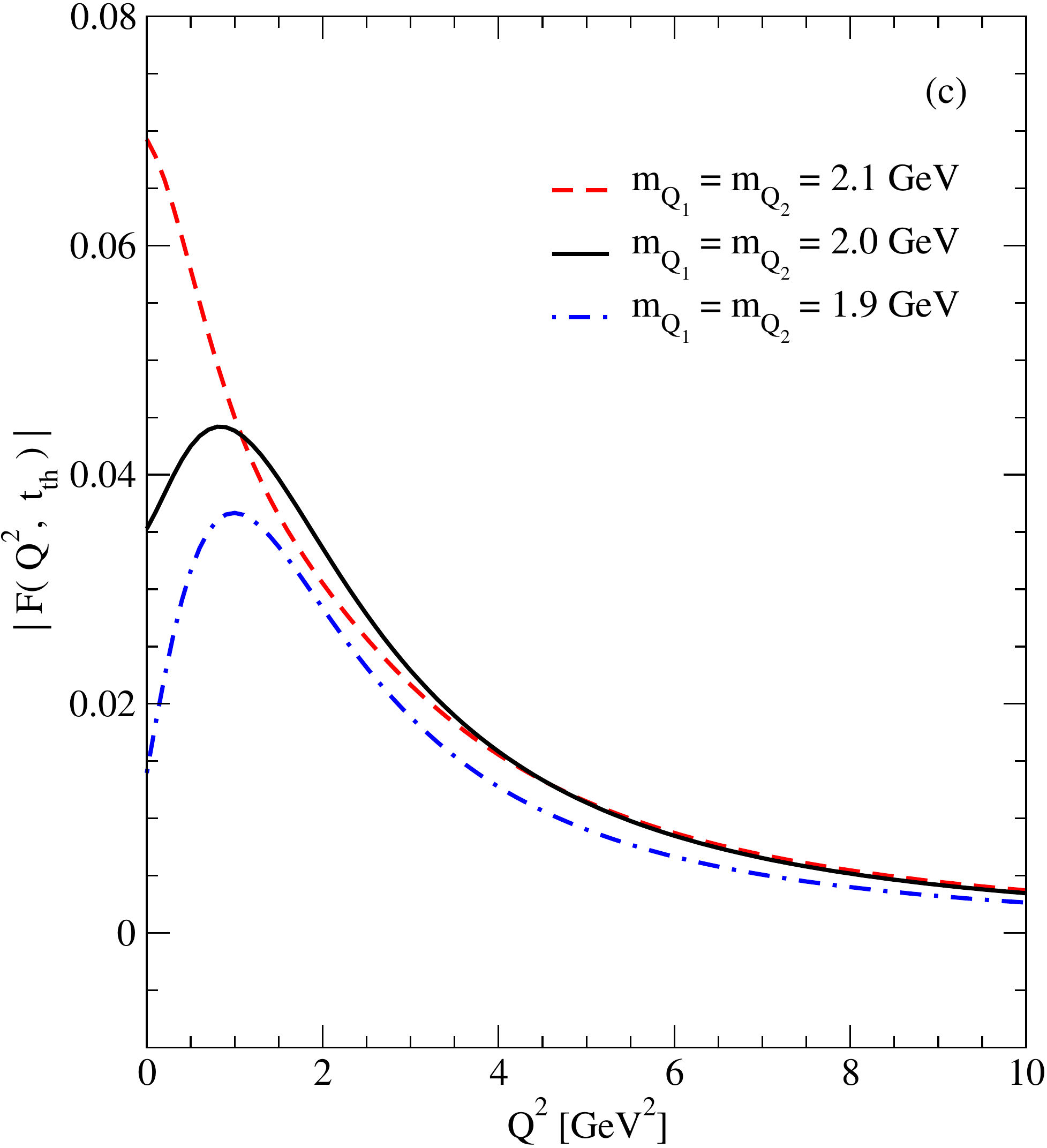}
\caption{\label{fig6}%
The Compton form factor $\mathcal{F} ^{\rm VMP}_{c}(Q^{2}, t)$. 
(a) its real part, (b), its imaginary part, and (c) its modulus for three constituent masses, 
$m_{Q_{1}}=m_{Q_{2}}=1.9$, $2.0$, and $2.1$~GeV at the momentum transfer squared 
$t=t_{\rm th} \simeq-0.7593~\mbox{GeV}^{2}$.}
\end{figure*}

We then compute the CFF in VMP, $\mathcal{F} ^{\rm VMP}_{c}(Q^2, t)$. Shown in Fig.~\ref{fig6} are the real part, the imaginary part, and the modulus of the amplitude 
$\mathcal{F} ^{\rm VMP}_{c}$ obtained from $\mathscr{M}^+_{\rm charged}$ of Eq.~(\ref{eq28c}) in the range of 
$0 \leq Q^2\leq 10$~GeV$^2$ at $t=t_{\rm th}(Q^2=0) \simeq -0.7593~\mbox{GeV}^2$. 
In order to explore the sensitivity of $\mathcal{F} ^{\rm VMP}_{c}(Q^2, t)$ on the constituent mass, we vary 
the constituent mass and repeat the computations for $m_{Q_1}=1.9$, $2.0$, and $2.1$~GeV, while keeping 
$m_{Q_1}=m_{Q_2}$.%
\footnote{These masses give the binding energy $B=m_{Q_{1}}+m_{Q_{2}}-M_{T}$ in the range of (0.1--0.5)~GeV.}
The results for $m_{Q_1}=1.9$, $2.0$, and $2.1$~GeV are presented by the dot-dashed, solid, and dashed 
lines, respectively.
The close inspection of Fig.~\ref{fig6} leads to the following comments.
(i) The real part has a hump structure, and the peak locates at the higher values of $Q^2$ with the lesser pronounced 
hump structures as the binding energy increases, as shown in Fig.~\ref{fig6}(a).
(ii) The magnitude of the imaginary part gets larger as the binding energy increases as shown in Fig.~\ref{fig6}(b).
(iii) As a result, the hump behavior of $\abs{\mathcal{F} ^{\rm VMP}_{c}(Q^2, t)}$ shown in Fig.~\ref{fig6}(c) 
appears strong near $Q^2 \simeq \abs{t_{\rm th}(Q^2=0)}$ for weak binding energies, but it goes away as the
binding energy increases. 
Also, there is no hump structure in $\abs{\mathcal{F} ^{\rm VMP}_{c}(Q^2, t)}$ for $Q^2 > \abs{t_{\rm th}(Q^2=0)}$ 
region and the binding energy effect is getting smaller as $Q^2$ increases.

\begin{figure}[t!]\centering
\includegraphics[width=4.2cm, height=4cm]{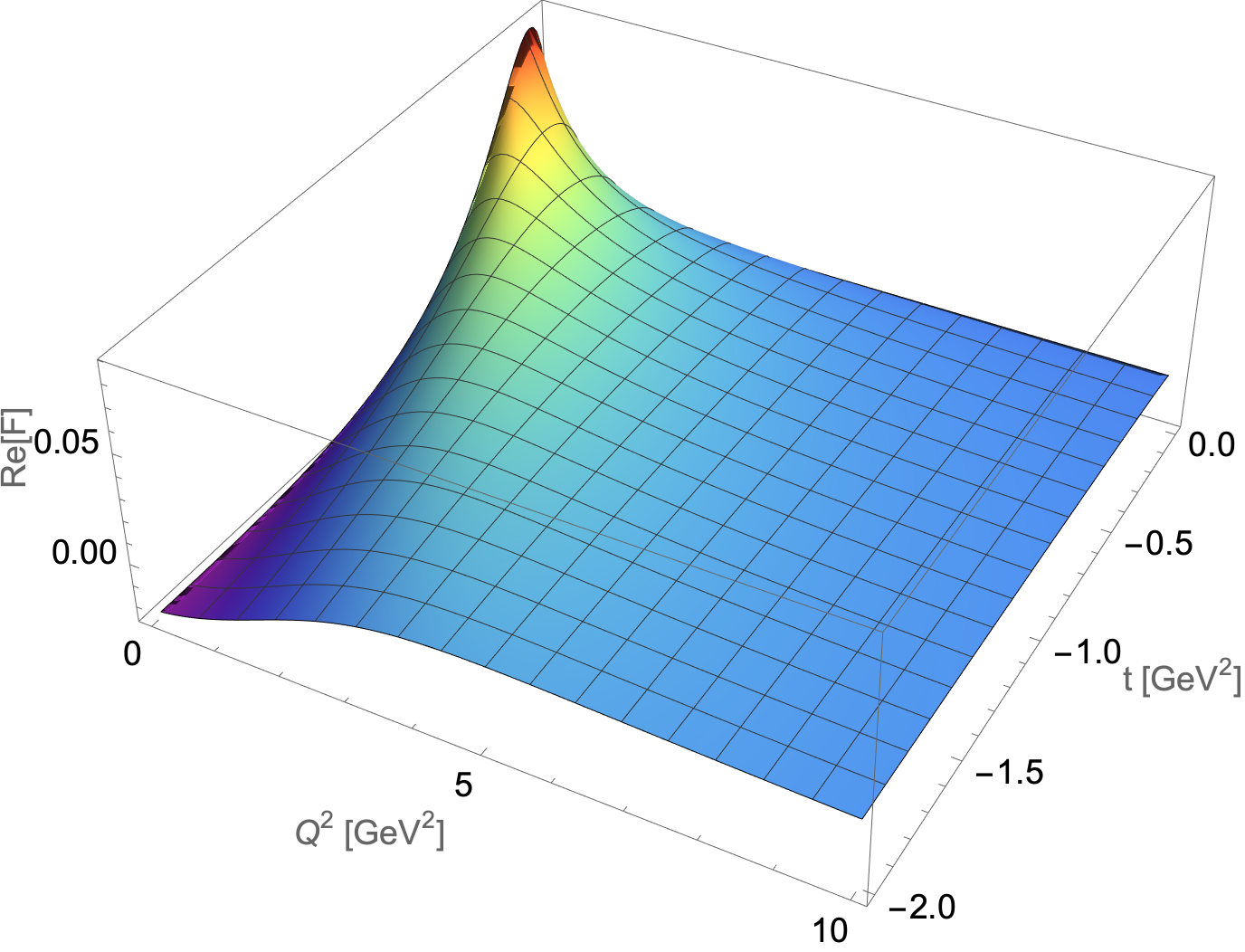}
\includegraphics[width=4.3cm, height=4cm]{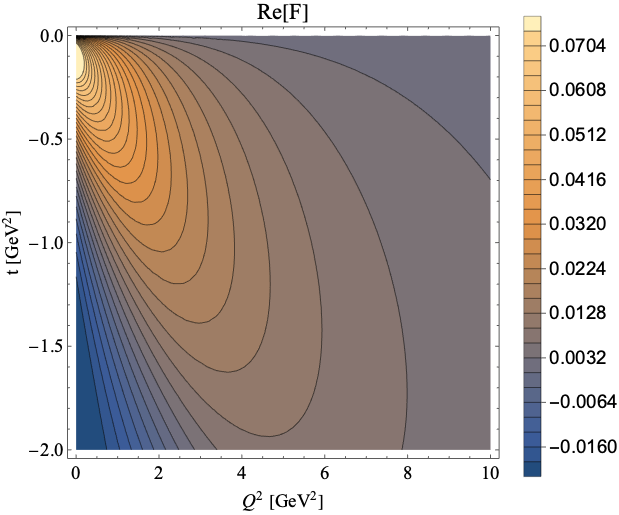}
\includegraphics[width=4.2cm, height=4cm]{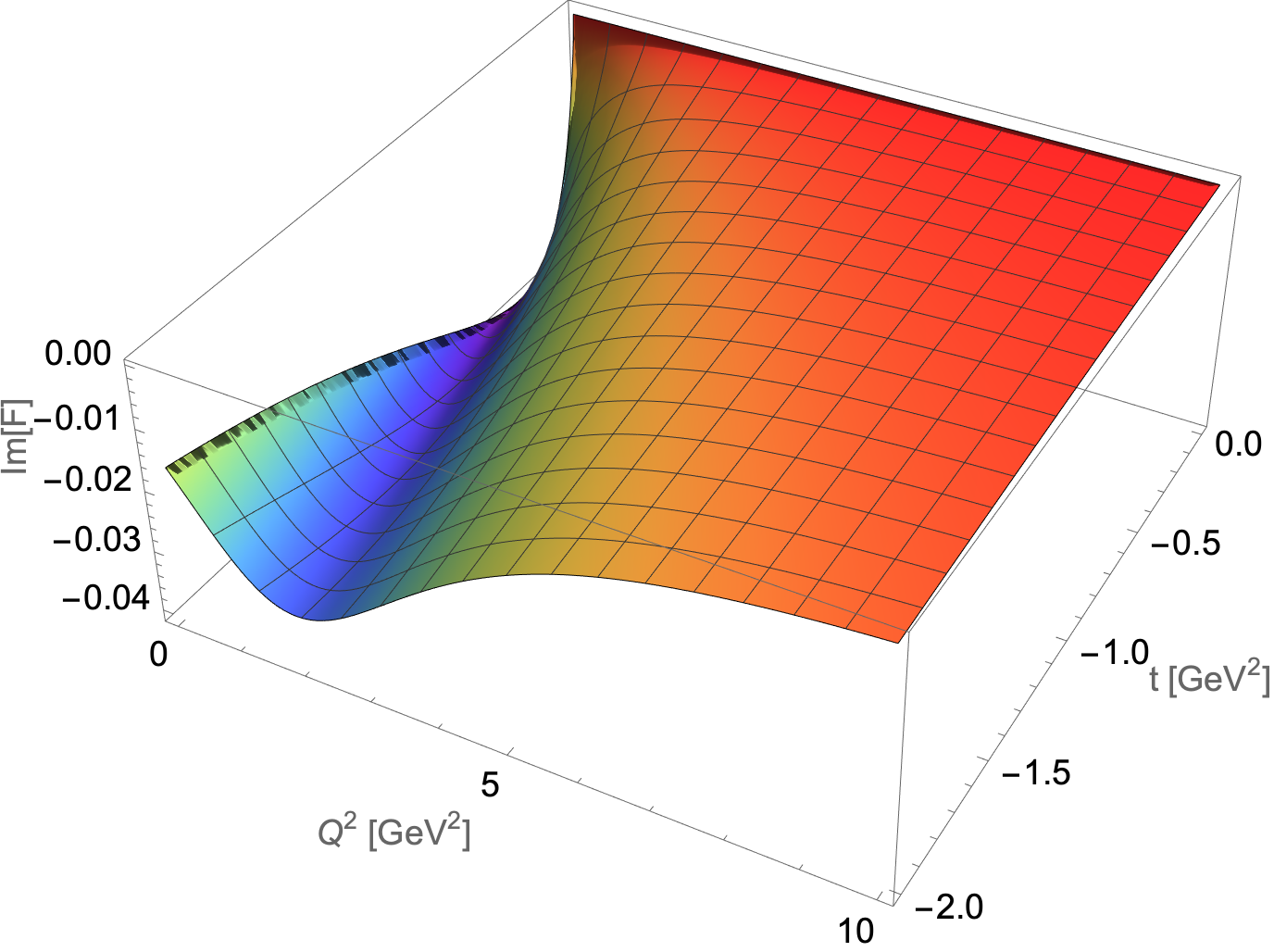}
\includegraphics[width=4.3cm, height=4cm]{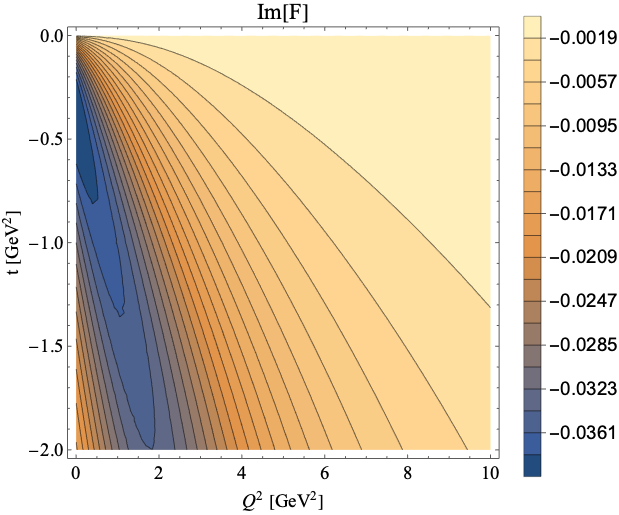}
\includegraphics[width=4.2cm, height=4cm]{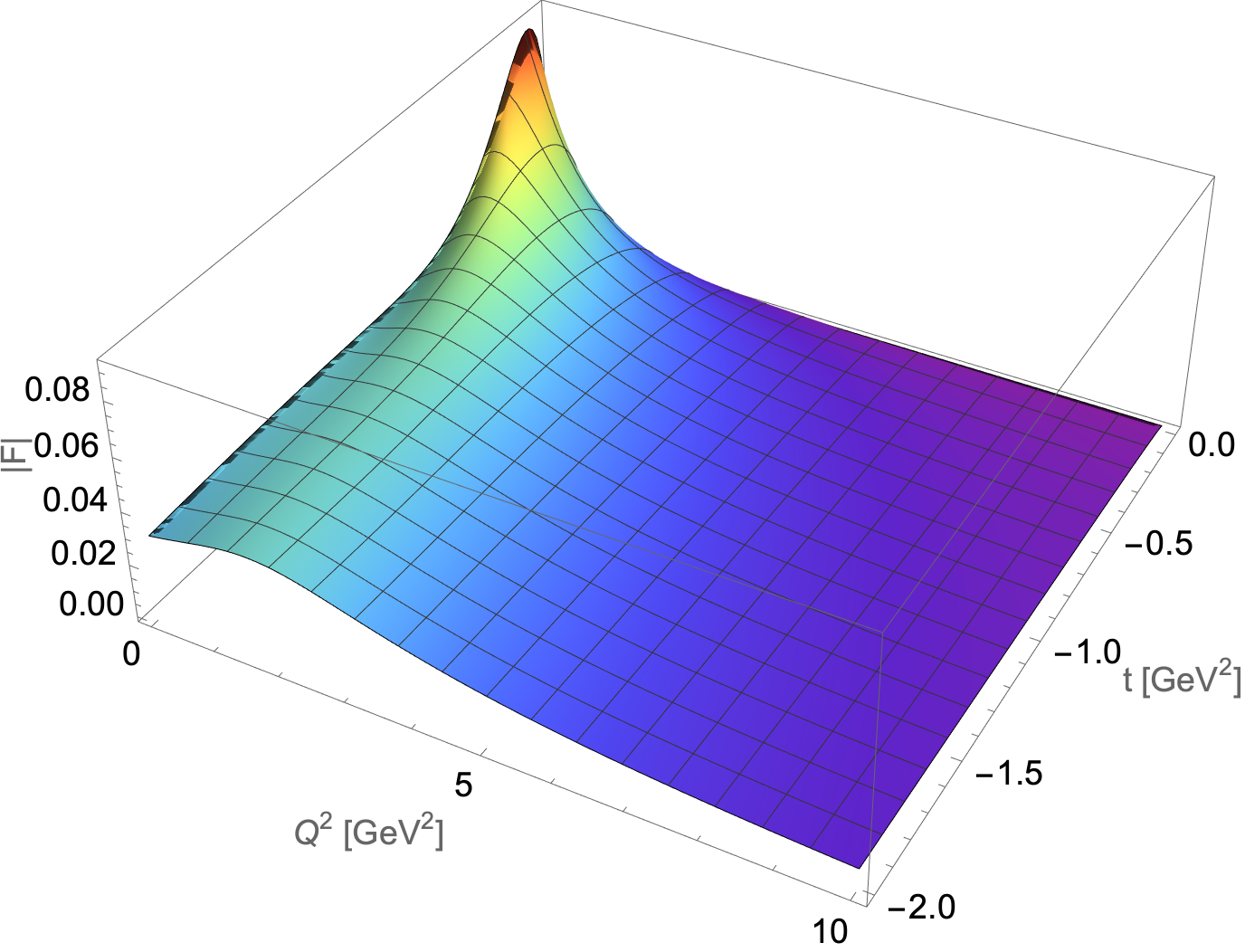}
\includegraphics[width=4.3cm, height=4cm]{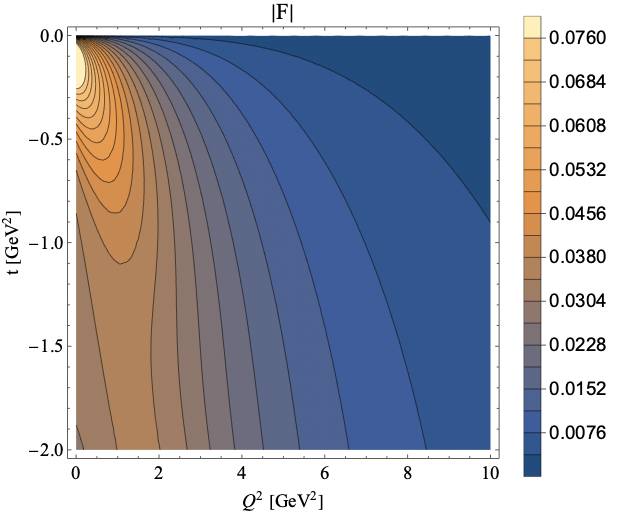}
\caption{\label{fig7}
(Left) Three-dimensional and (right) contour plots for the real part, imaginary part, and modulus of the 
Compton form factor $\mathcal{F} ^{\rm VMP}_{c}$ at $t \simeq -0.7593~\mbox{GeV}^{2}$.
}
\end{figure}

The left and right panels of Fig.~\ref{fig7} respectively show the three-dimensional and contour plots of 
$\mbox{Re} [\mathcal{F} ^{\rm VMP}_{c}]$, $\mbox{Im} [ \mathcal{F} ^{\rm VMP}_{c} ]$, 
and $\abs{\mathcal{F} ^{\rm VMP}_{c}}$ for the range of $0 \leq Q^{2} \leq 20~\mbox{GeV}^{2}$ and 
$-2~\mbox{GeV}^2 \leq t \leq 0$. 
Both the real (top panel) and imaginary (middle panel) parts are going to zero as $t\rightarrow 0$ 
regardless of the value of $Q^{2}$. 
For $\abs{t} \lesssim Q^2$, the real part of $\mathcal{F} ^{\rm VMP}_{c}$ shows a gradual crest along the straight line 
of $\abs{t} \approx Q^{2}$, and the imaginary part has a trough located at $\abs{t} \approx Q^{2}$. 
Both the real and imaginary parts rapidly approach to zero as $\abs{t}$ decreases to zero for the small $Q^{2}$ region 
from the crest and trough, respectively. 
The modulus of the CFF also has a crest around $\abs{t} \approx Q^{2}$ and gradually decreases as $Q^{2}$ increases 
and $\abs{t}$ decreases.

\subsection{CFF and GPD in DVMP limit}
\label{sec:5-B}

\begin{figure}[t!]\centering
\includegraphics[width=4.2cm, height=4cm]{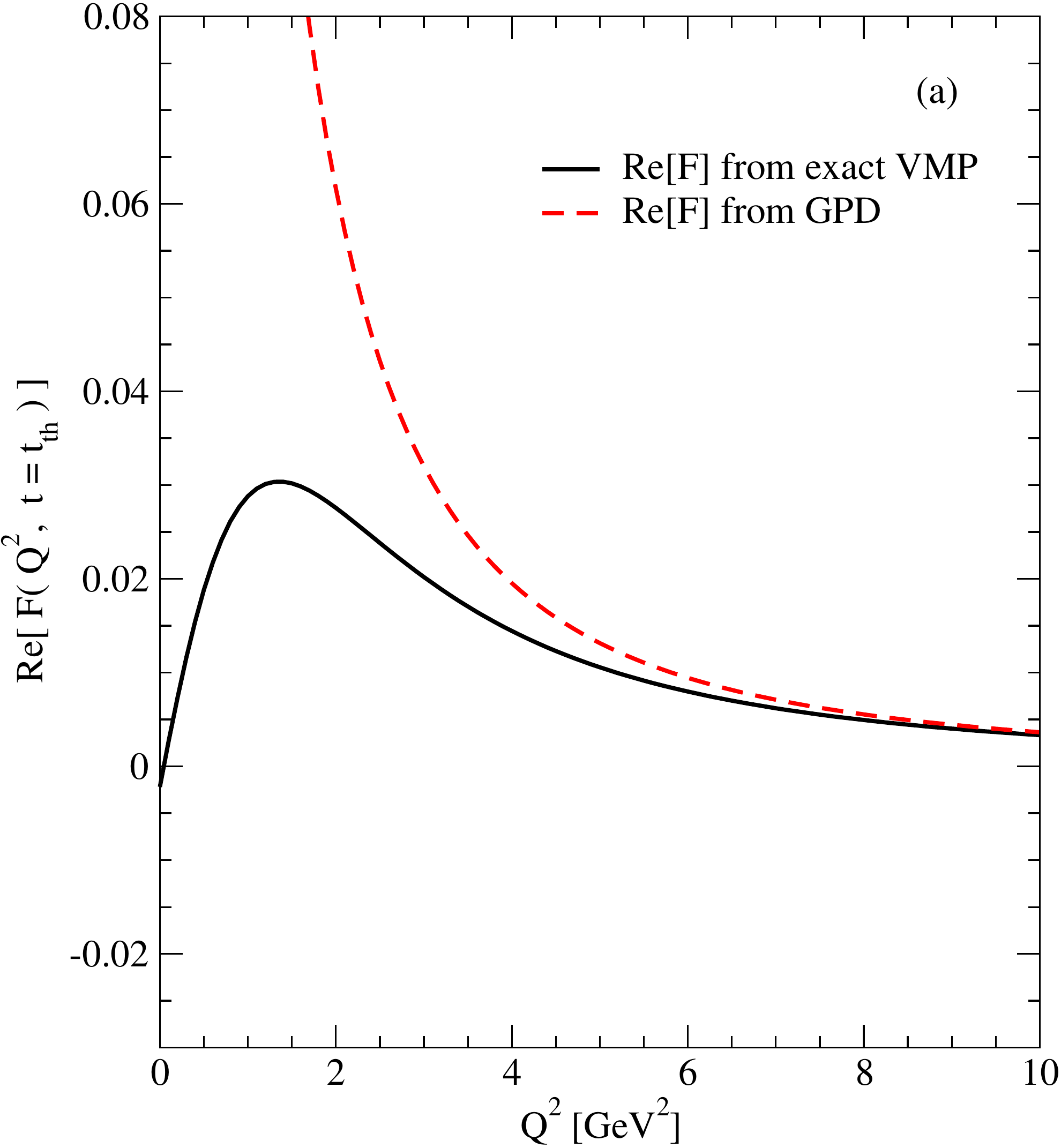}
\includegraphics[width=4.3cm, height=4cm]{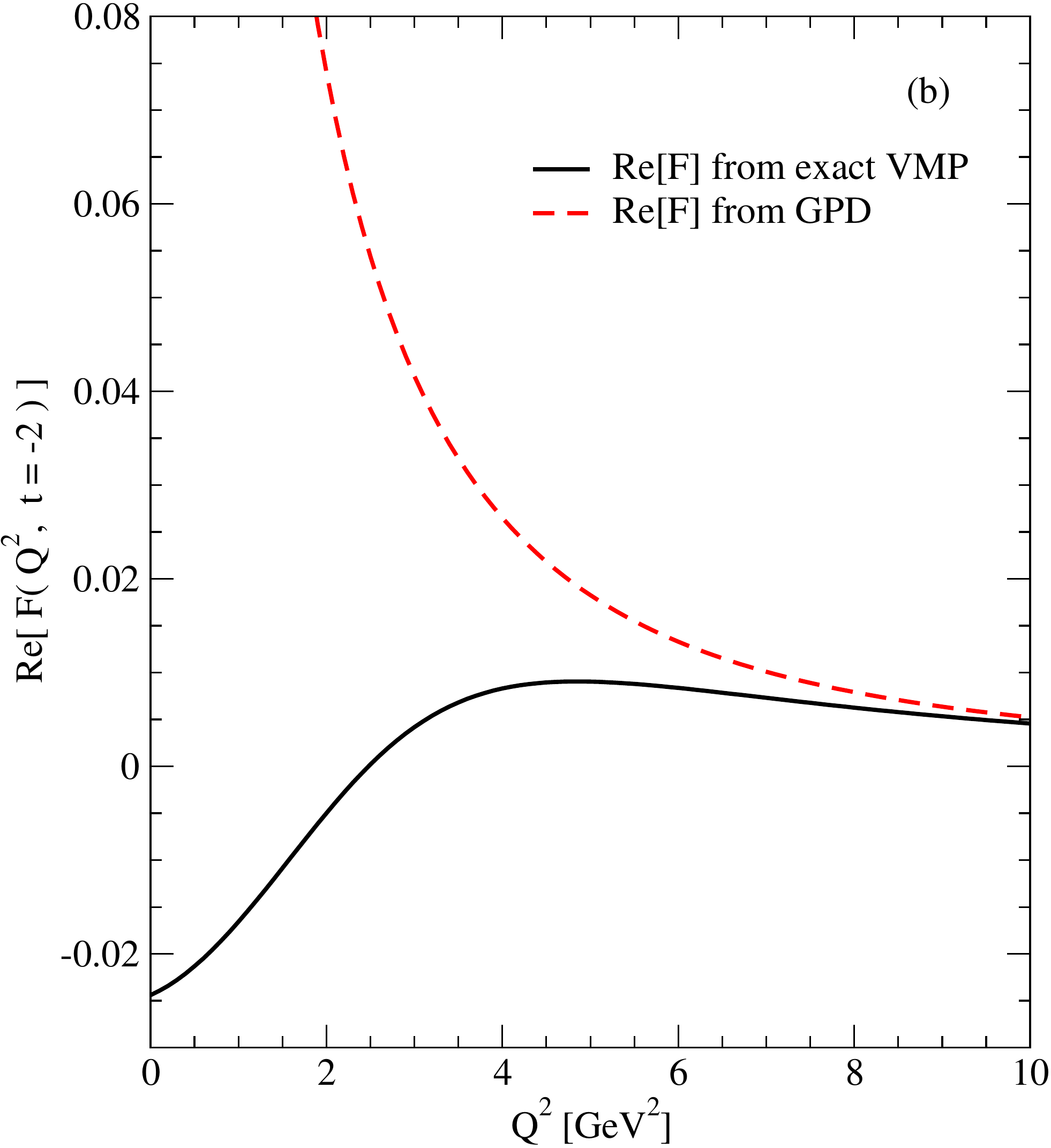}
\includegraphics[width=4.2cm, height=4cm]{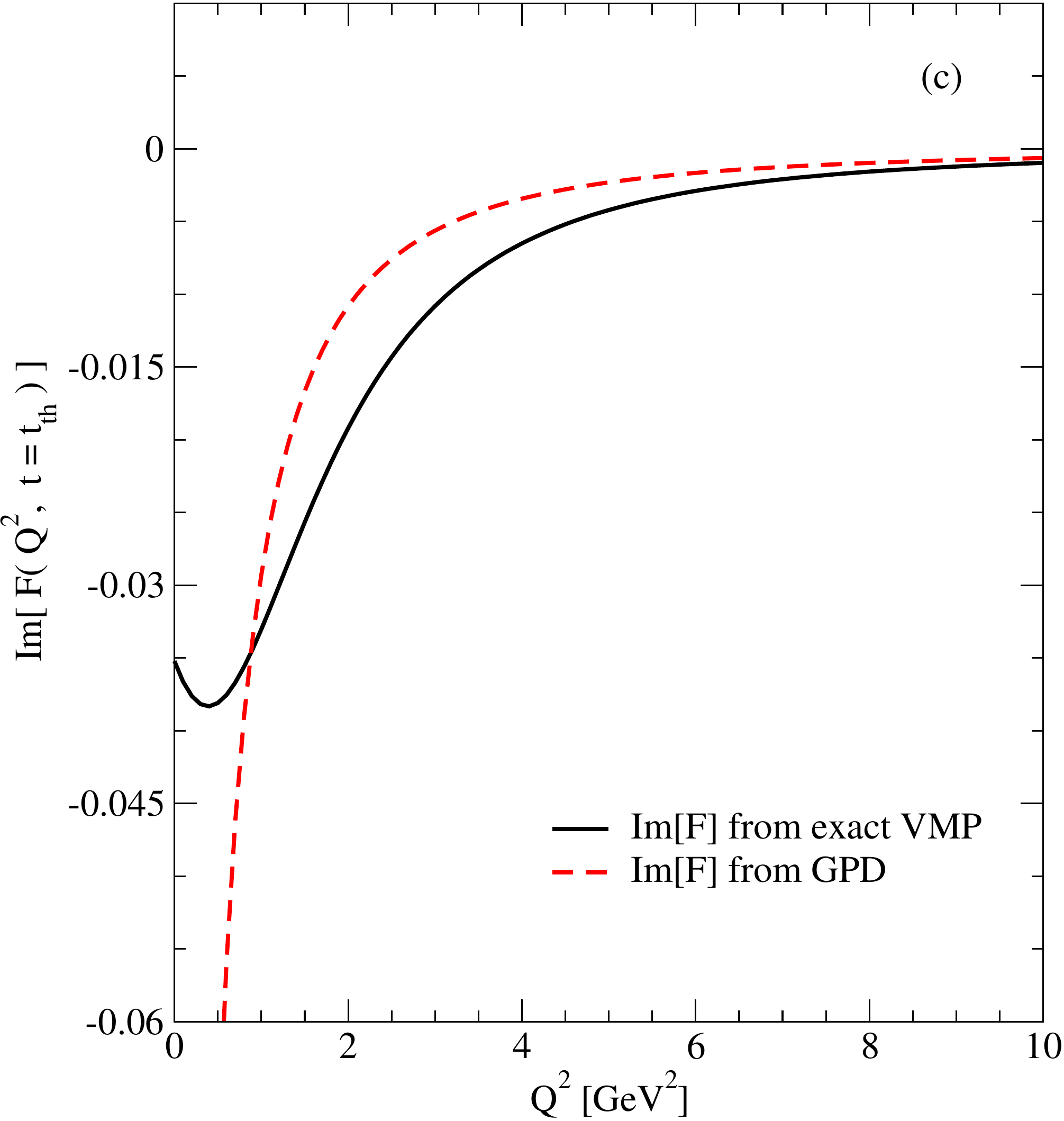}
\includegraphics[width=4.3cm, height=4cm]{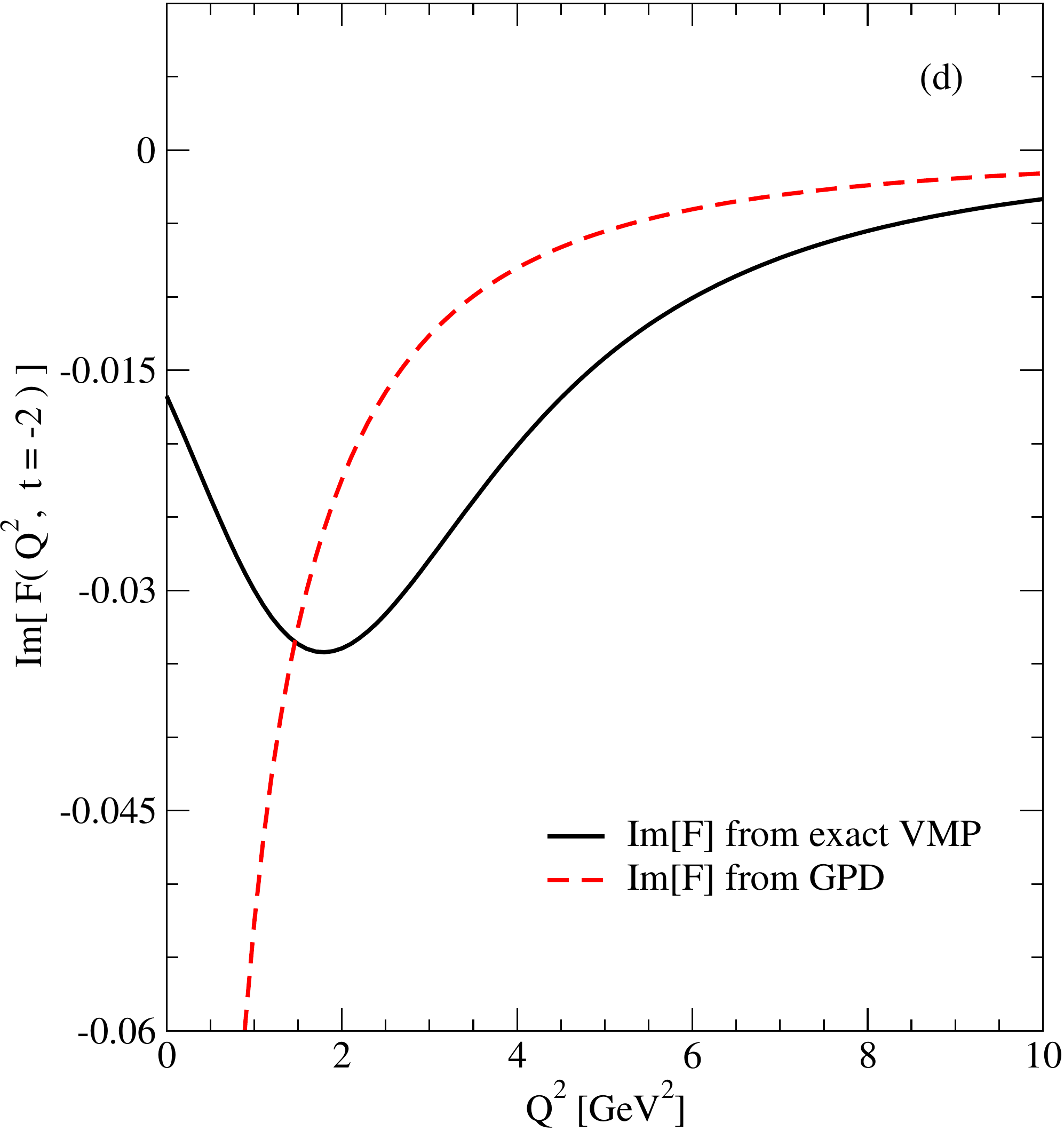}
\caption{\label{fig8}
The real and imaginary parts of the Compton form factor determined from $\mathscr{M}^{+}_{s+u+c+\rm ET}$ and 
$\mathscr{M}^{+\,\rm DVMP}$ (left panel) for $t \simeq- 0.7593$~GeV$^{2}$ and (right panel) for $t=-2$~GeV$^{2}$.
}
\end{figure}

In the DVMP limit, where $Q^2 \gg (M^2_S, \abs{t})$ but not 
explicitly involving $Q^2 \gg M^2_T$ due to the correlation among $\zeta_{1+1}, t$ and $M^2_T$ 
as discussed in Sec.~\ref{sec:2}, the scattering amplitude for the scalar meson production from either neutral or
charged scalar target is reduced to the DVMP amplitude $\mathscr{M}^{+\, \rm DVMP}_{s+u}$. 
Regardless of neutral or charged target, the 
$s$- and $u$-channel amplitudes are factorized in the DVMP limit
as discussed in Sec.~\ref{sec:4-A} and $\mathscr{M}^{+\, \rm DVMP}_{s+u}$ is given by the factorized form of the hard scattering part and the soft GPD $H(\zeta, x,t)$ as shown in Eq.(\ref{eq31c}).
In order to find the region where the DVMP limit is valid, we compare the CFF ${\mathcal F}^{\rm DVMP}$ 
in Eq.~(\ref{eq36c}) obtained from $\mathscr{M}^{+\rm DVMP}_{s+u}$ with the exact solution 
${\mathcal F} ^{\rm VMP}_{c}$ presented in Fig.~\ref{fig6}.

In Fig.~\ref{fig8}, we compare ${\mathcal F} ^{\rm VMP}_{c}$ (solid lines) and the leading twist 
${\mathcal F}^{\rm DVMP}$ (dashed lines) for the range of $0 \leq Q^2 \leq 10$~GeV$^2$.
Figures~\ref{fig8} (a,c) and (b,d) show ($\mbox{Re} [ \mathcal{F} ]$, $\mbox{Im} [ \mathcal{F} ]$), 
at $t=t_{\rm th}(Q^2=0) \simeq -0.7593$~GeV$^{2}$ and $t= -2$~GeV$^{2}$, respectively.
There are several points to comment on the results shown in Fig.~\ref{fig8}.
(i) While the exact solution ${\mathcal F} ^{\rm VMP}_{c}$ is finite at $Q^2=0$ with a hump behavior near $Q^2 = -t$, 
the results of ${\mathcal F}^{\rm DVMP}$ obtained at the leading order 
of $Q^2$ in the DVMP limit do not have any hump structure but blows up in the vicinity of $Q^{2}=0$. 
(ii) The agreement between ${\mathcal F} ^{\rm VMP}_{c}$ and ${\mathcal F}^{\rm DVMP}$ can be seen 
at large $Q^2$ region, but it reaches faster as the smaller $-t$ value is used.
For instance, the VMP and DVMP results agree for $Q^{2} \geq 6~{\rm GeV^{2}}$ when $-t \simeq 0.7593~\mbox{GeV}^{2}$ 
is fixed. 
This indicates that the validity of the GPD handbag approximation is governed not just either by $-t$ or $Q^2$, but by $-t/Q^2$.
Figure~\ref{fig8} shows that it is valid in the region of $-t/Q^2 \leq 0.125$ with $-t \simeq 0.7593~\mbox{GeV}^{2}$.
On the other hand, for $-t \simeq 2~\mbox{GeV}^{2}$, the agreement of the VMP and DVMP CFFs can be seen at
higher $Q^2$, i.e., $Q^2 > 10$~GeV$^2$, which corresponds to $-t/Q^2 \simeq 0.2$.
To have better agreement both for the real and imaginary parts of the CFF, $Q^2$ should be even larger to get 
$-t/Q^2 \lesssim 0.1$.
Therefore, we find that the GPD handbag approximation can be valid only for a small value of $-t/Q^2$,
although the critical value of $-t/Q^2$ appears somewhat larger as $-t$ increases in our ($1+1$) dimensional results.
This indicates that, for realistic VMP measurements in $(3+1)$ dimensions, very forward scattering region should 
be required to invoke the GPD handbag approximation as the forward scattering region would allow a very small value of $-t$.
Since $-t$ is not independent of the target mass $M_T$ as shown in Eq.~(\ref{eq4c}) in ($1+1$) dimensions, the 
skewness parameter $\zeta$ gets smaller as $M_T$ increases for given $-t$ value.
This indicates that the GPD handbag approximation agrees with the exact VMP result faster with a larger $M_T$ 
than a smaller $M_T$, which appears the characteristic of the $(1+1)$ dimensional analysis.

\begin{figure}[t!]\centering
\includegraphics[width=\columnwidth]{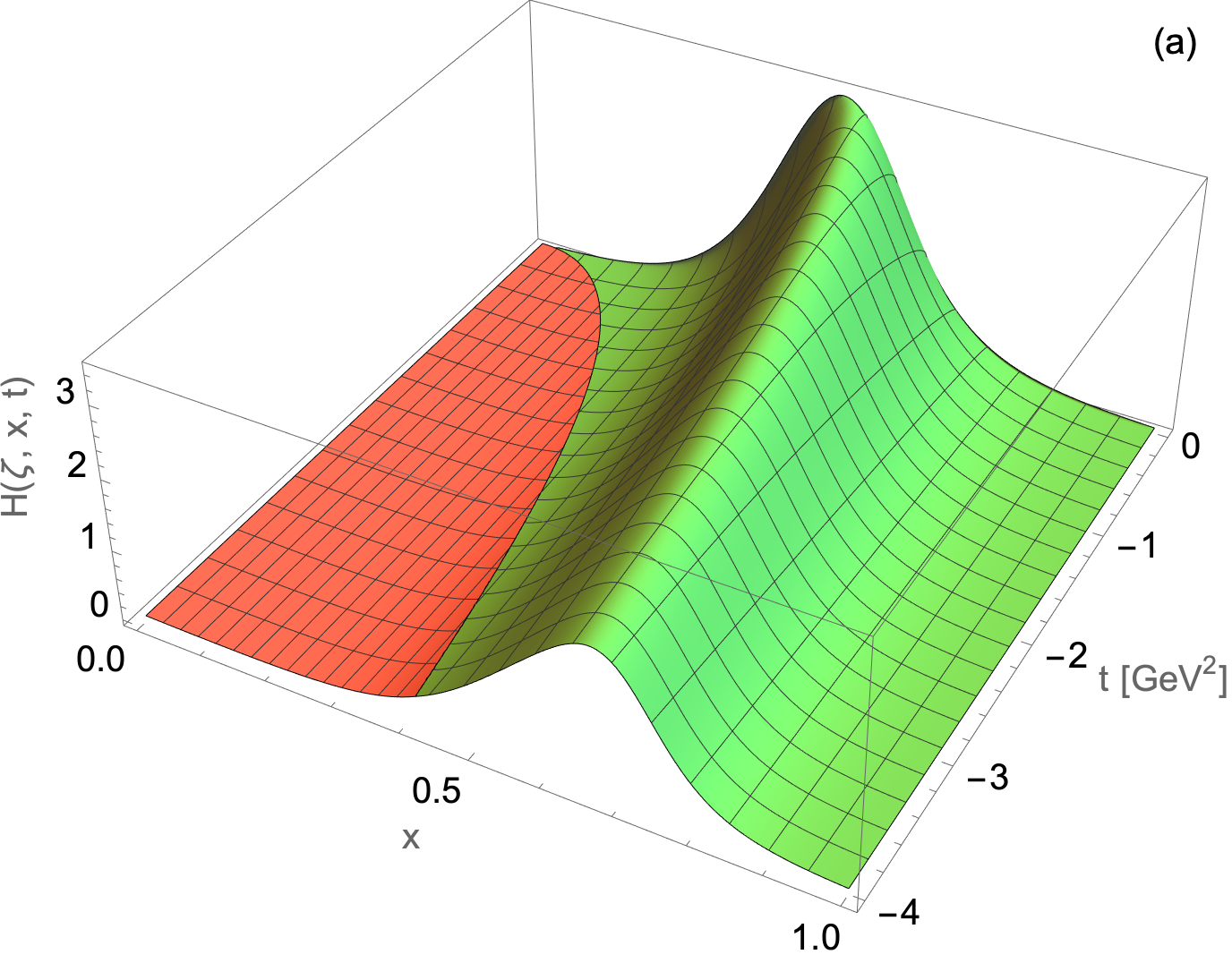}
\includegraphics[width=\columnwidth]{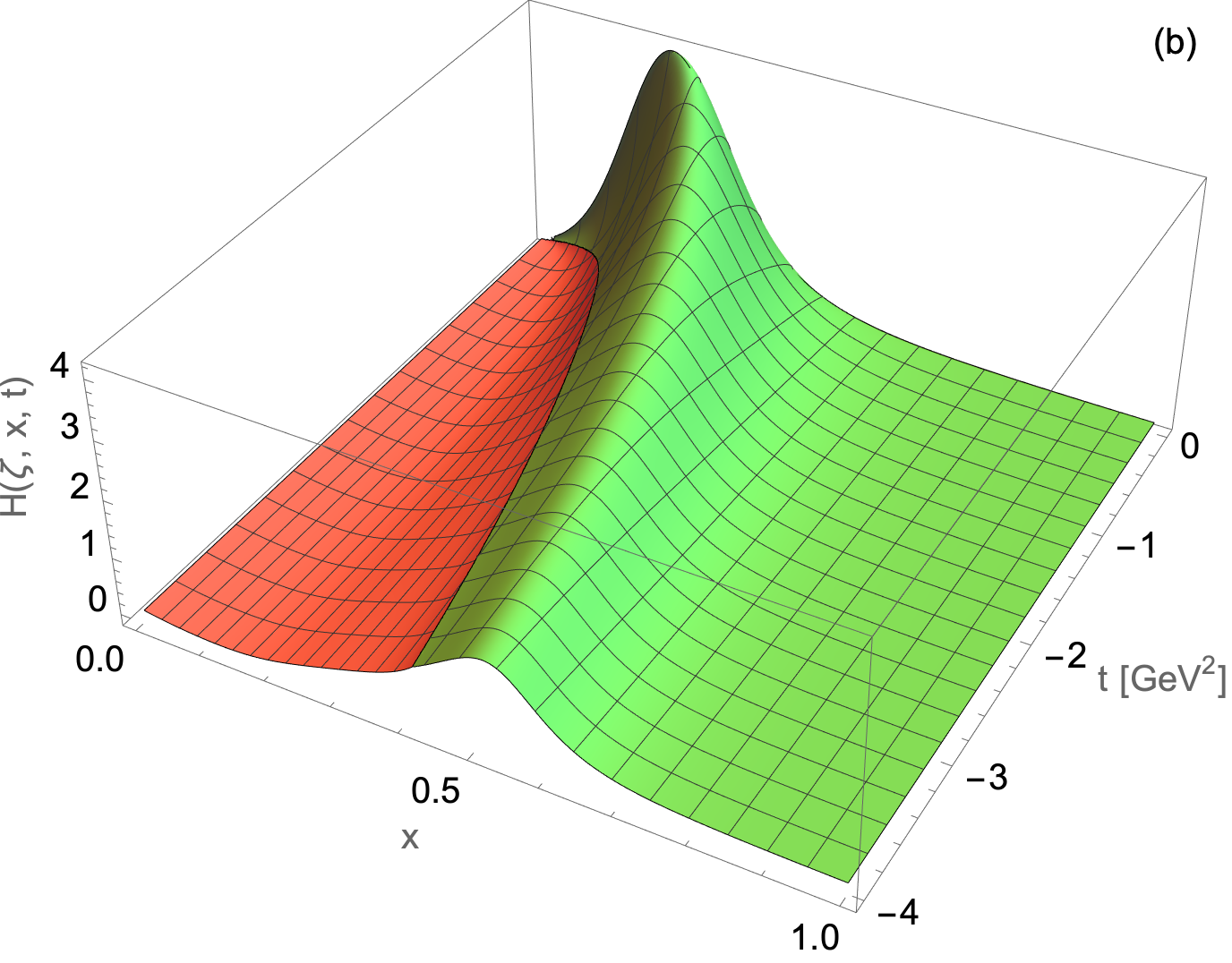}
\caption{\label{fig9}
Three-dimensional plots of $H(\zeta, x, t)$ for two parameter sets. (Upper panel) $(m_{Q_1}, m_{Q_2})=(2,2) $~GeV and
(lower panel) $(m_{Q_1}, m_{Q_2})=(1, 3)$~GeV, in the region of $0 \leq x \leq 1$ and $-4~\mbox{GeV}^2 \leq t \leq 0$.}
\end{figure}

As we have mentioned before, in $(1+1)$ dimensions, the GPD $H(\zeta, x,t)$ given by Eq.~(\ref{eq33c}) is 
essentially a function of $x$ and $t$ since $\zeta$ and $t$ are related to each other by Eq.~(\ref{eq5c}).
Figure~\ref{fig9} shows the three-dimensional plots of $H(\zeta, x, t)$ for two parameter sets, 
$(m_{Q_1}, m_{Q_2})=(2,2) $~GeV and $(m_{Q_1}, m_{Q_2})=(1, 3)$~GeV, of which results are presented in the 
upper and lower panels of Fig.~\ref{fig9}, respectively, in the range of $0 \leq x \leq 1$ and 
$-4~\mbox{GeV}^2 \leq t \leq 0$. 
The red and green regions correspond to GPDs in the ERBL ($0 \leq x \leq \zeta$) and DGLAP 
($\zeta \leq x \leq 1$) regions, respectively.
The crossover boundaries (black lines) between the two regions correspond to the lines $x=\zeta$, i.e.,
$H_{\rm ERBL}(\zeta,\zeta,t)=H_{\rm DGLAP}(\zeta,\zeta,t)$.
Again $t$ and $\zeta$ are not independent variables in $(1+1)$ dimensions. In the crossover boundary for a given parameter set, the longitudinal momentum fraction $x$ carried by the struck constituent $Q_1$ 
gradually increases as $\abs{t}$ increases. 
Also, the peak position of GPD always exists in the DGLAP region. 
Comparing the two parameter sets, the value of $x$ at the peak is found to decrease as the mass ratio 
$m_{Q_1}/m_{Q_2}$ decreases.

\subsection{PDF and EM form factor}
\label{sec:5-C}

\begin{figure}[t!]\centering
\includegraphics[width=\columnwidth]{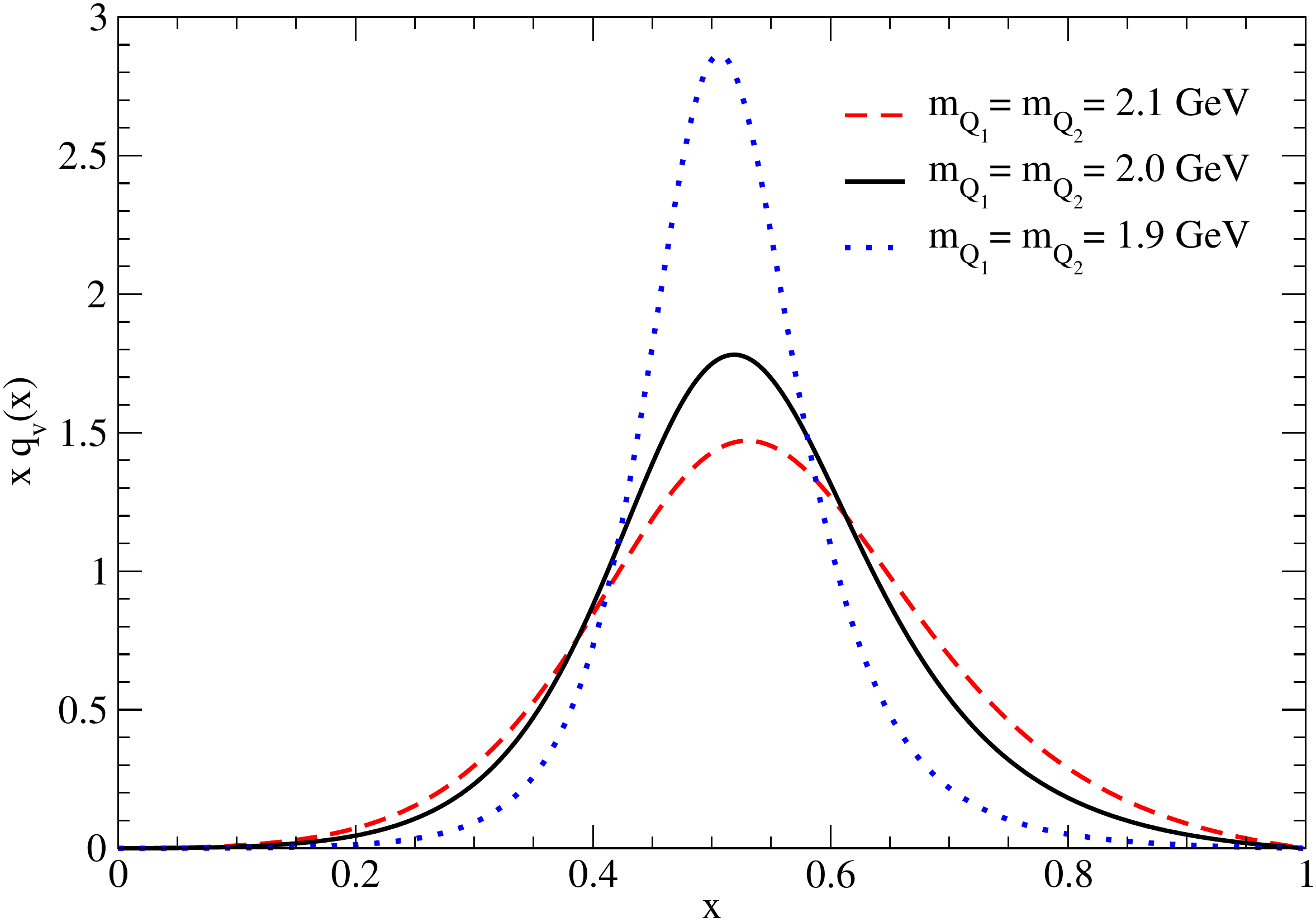}
\caption{\label{fig10}
The ordinary valence PDF $q_v(x)$ of the scalar target multiplied by $x$ for $m_{Q_1}=m_{Q_2}= 1.9$, $2.0$, 
and $2.1$~GeV.}
\end{figure}

Figure~\ref{fig10} shows the ordinary valence PDF for the ``helium'' target multiplied by $x$, i.e., $x q_v(x)$ 
for three values of $m_{Q_1}=m_{Q_2}$ with $M_T = 3.7$~GeV.
The dotted, solid, and dashed lines in this figure represent the results obtained with $m_{Q_{1(2)}}=1.9$, 
$2.0$, and $2.1$~GeV, respectively.
Since $q_v(x)$ in Eq.~(\ref{eq38c}) is symmetric under the exchange of $x \to 1-x$ for the case of equal 
constituent mass, $x q_v(x)$ is somehow asymmetric. 
In Fig.~\ref{fig10}, this asymmetric behavior is getting noticeable as the binding gets stronger.

\begin{table}[t!]
\centering
\caption{The \textit{n}-th moments $\langle y_{n} \rangle$ of the parton distribution function for the 
scalar target with three constituent mass sets.}
\label{tab1}
\renewcommand{\arraystretch}{1.2}
\setlength{\tabcolsep}{5.5pt}
\begin{tabular}{ccccccc} \hline\hline
$m_{Q_{1}}=m_{Q_{2}}$ & $\langle y_1 \rangle$ & $\langle y_2 \rangle$ & $\langle y_3 \rangle$  & 
$\langle y_4 \rangle$ & $\langle y_5 \rangle$ & $\langle y_6 \rangle$\\
\hline
$2.1$ GeV  &  0  &  0.0884 & 0 & 0.0242 & 0 & 0.0104\\
$2.0$ GeV  &  0  &  0.0662 & 0 & 0.0155 & 0 & 0.0061 \\
$1.9$ GeV  &  0  &  0.0313 & 0 & 0.0050 & 0 & 0.0016\\
\hline\hline
\end{tabular}
\end{table}

\begin{figure}[t!]\centering
\includegraphics[width=\columnwidth]{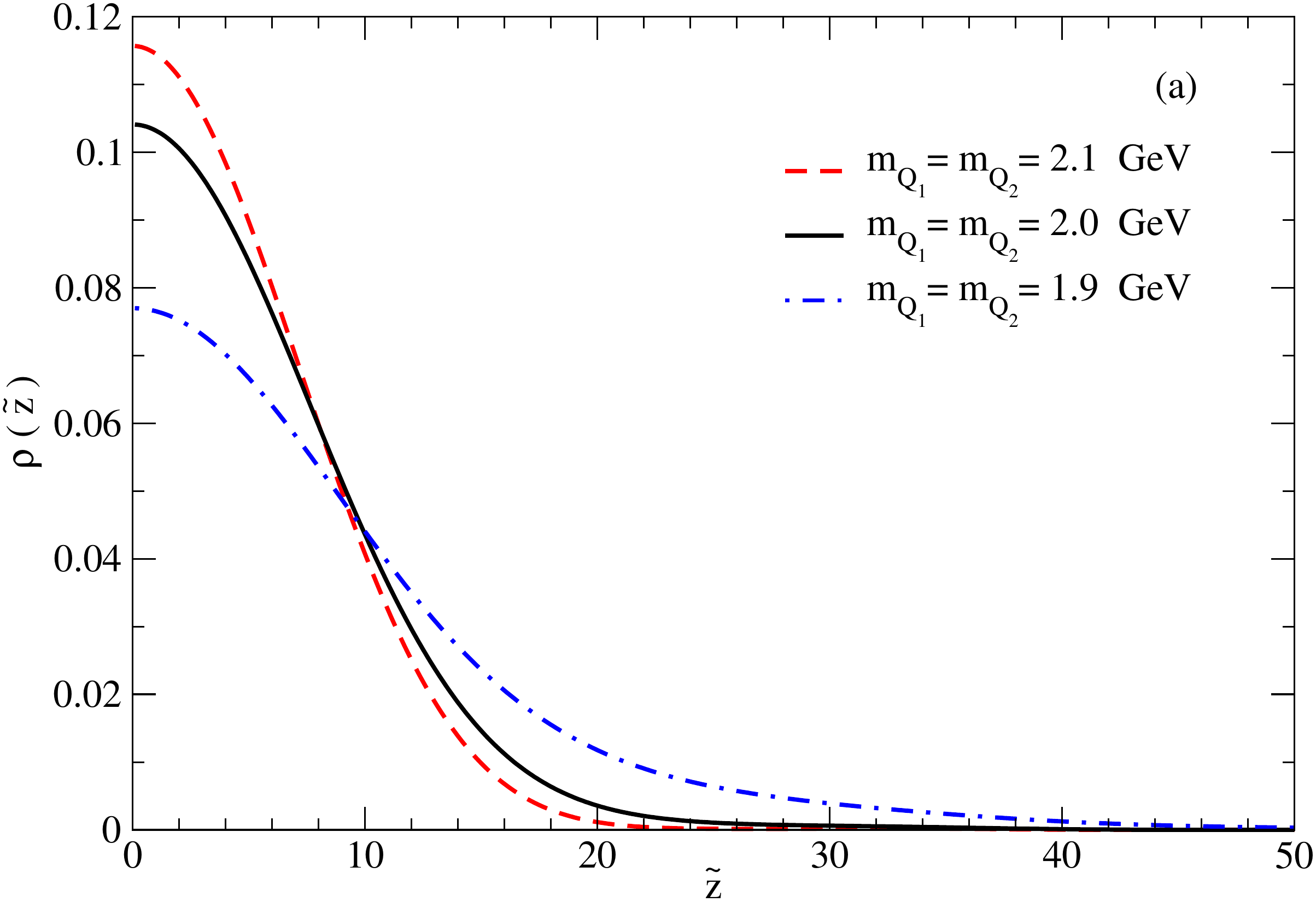}
\includegraphics[width=\columnwidth]{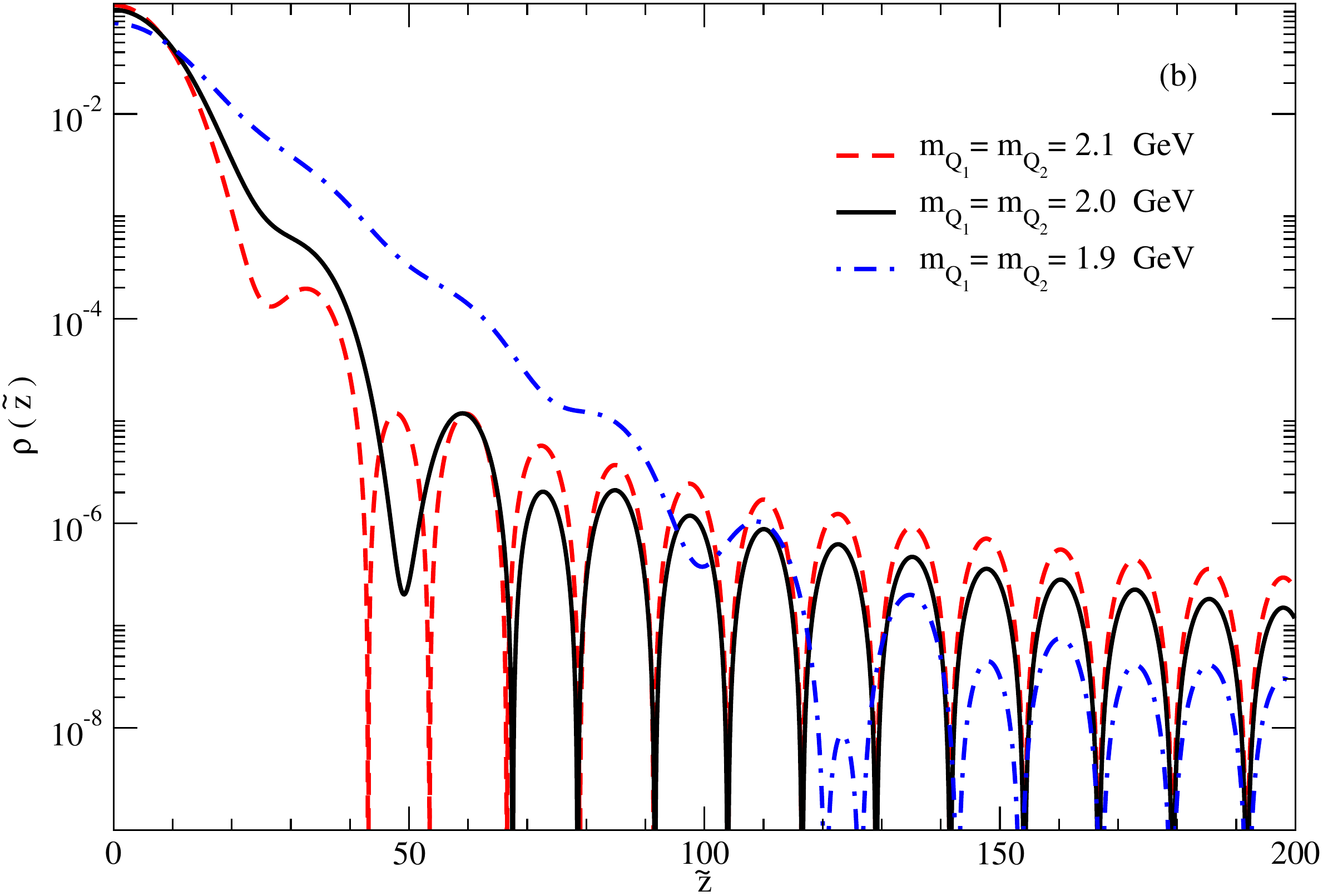}
\caption{\label{fig11}%
Longitudinal probability density $\varrho(\tilde{z})$ for the scalar target with $m_{Q_1}=m_{Q_2}=1.9$, 
$2.0$, and $2.1$~GeV in the LF coordinate space $\tilde{z}$. 
(a) $\varrho(\tilde{z})$ in linear scale for the range of $0 \leq \tilde{z} \leq 50$, and 
(b) $\varrho(\tilde{z})$ in logarithmic scale in the range of $0 \leq \tilde{z} \leq 200$.}
\end{figure}

The $n$-th moments of $q_v(x)$ for the scalar target are summarized in Table~\ref{tab1}. 
Since $q_v(y)$ in Eq.~(\ref{eq40c}) is an even function of $y$, the odd-numbered moments vanish. 
Our results in Table~\ref{tab1} show that the heavier the constituent mass (or equivalently, the larger the 
binding energy) is, the greater the values of even-numbered moments are.
This implies that the shape of the PDF, $q_v(y)$, is more narrowly peaked at $y=0$ and more suppressed at 
the endpoints $(y = \pm 1)$ as the binding energy of the scalar target decreases.

Shown in Fig.~\ref{fig11} is the longitudinal probability density $\varrho(\tilde{z})=|\psi(\tilde{z})|^2$ (See Eqs.(\ref{eq41c}) and (\ref{eq42c}))
for the scalar target with $M_T = 3.7$~GeV in the LF coordinate space of $\tilde{z}=x^- p^+$ which is 
completely Lorentz-invariant in $(1+1)$ dimensions. 
The dot-dashed, solid, and dashed lines represent the results for $m_{Q_{1(2)}}=1.9$, $2.0$, and $2.1$~GeV, 
respectively.
In order to clearly show the behavior of the longitudinal probability density, we plot $\varrho(\tilde{z})$ 
in two ways.
In Fig.~\ref{fig11}(a), $\varrho(\tilde{z})$ is shown in the range of $0 \leq \tilde{z} \leq 50$ in linear scale.
This shows that the stronger bound state ($m_{Q_{1(2)}}=2.1$~GeV) has a more concentrated distribution near 
$\tilde{z}=0$ than weakly bound states have.
The long range behavior of $\varrho(\tilde{z})$ is shown in Fig.~\ref{fig11}(b) that plots the same function 
in logarithmic scale for a wider range of $\tilde{z}$.
One can verify that $\varrho(\tilde{z})$ shows the oscillating behavior for large ${\tilde z}$. Furthermore,
the onset of the oscillation appears earlier, and the amplitude of the oscillation is larger for the strongly 
bound states than the weakly bound states. 
Our observation is consistent with those for the pion case reported in Refs.~\cite{MB19,BD07a,HLFHS18}.

\begin{figure}[t!]\centering
\includegraphics[width=0.8\columnwidth]{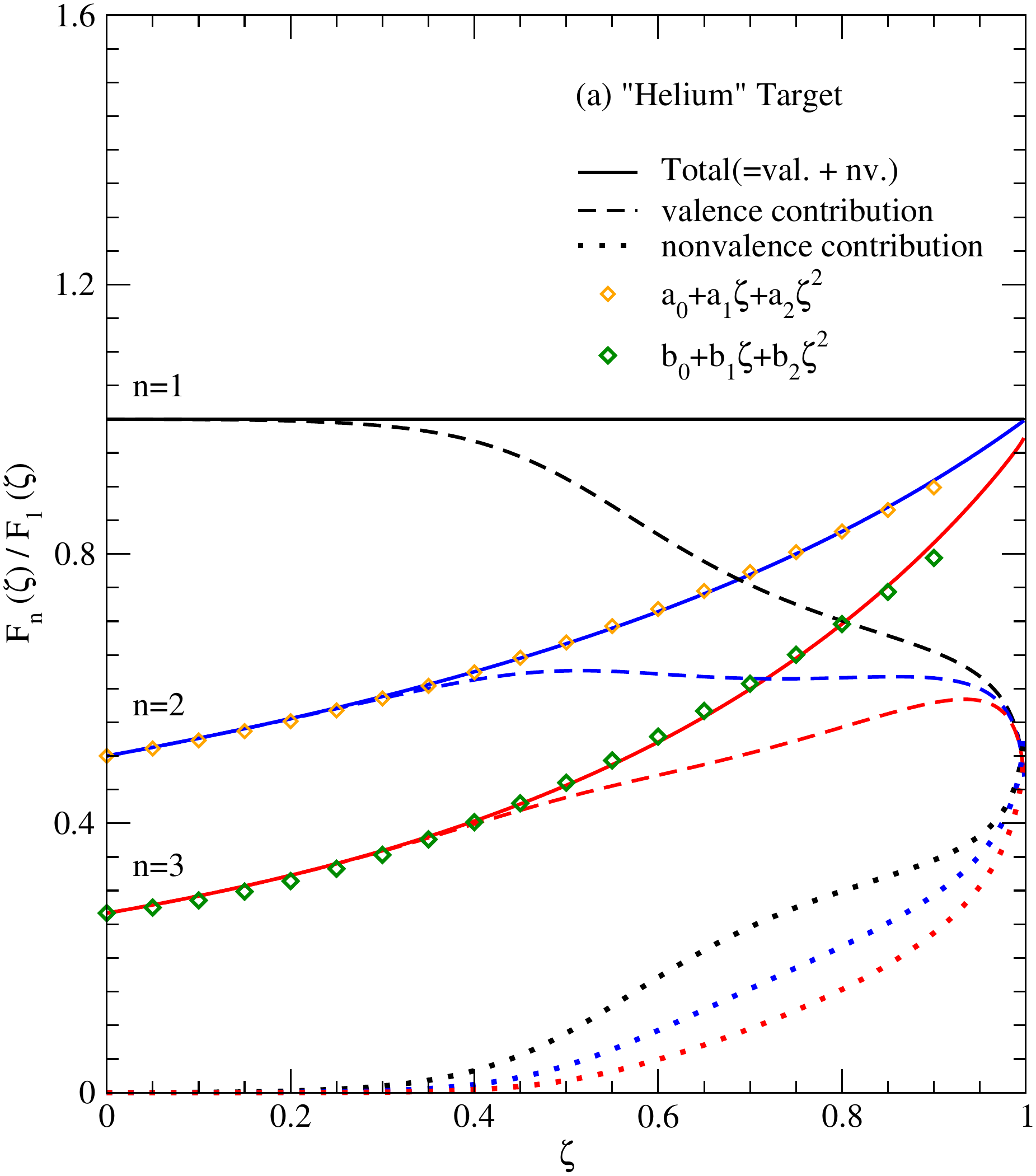} 

\bigskip\bigskip

\includegraphics[width=0.8\columnwidth]{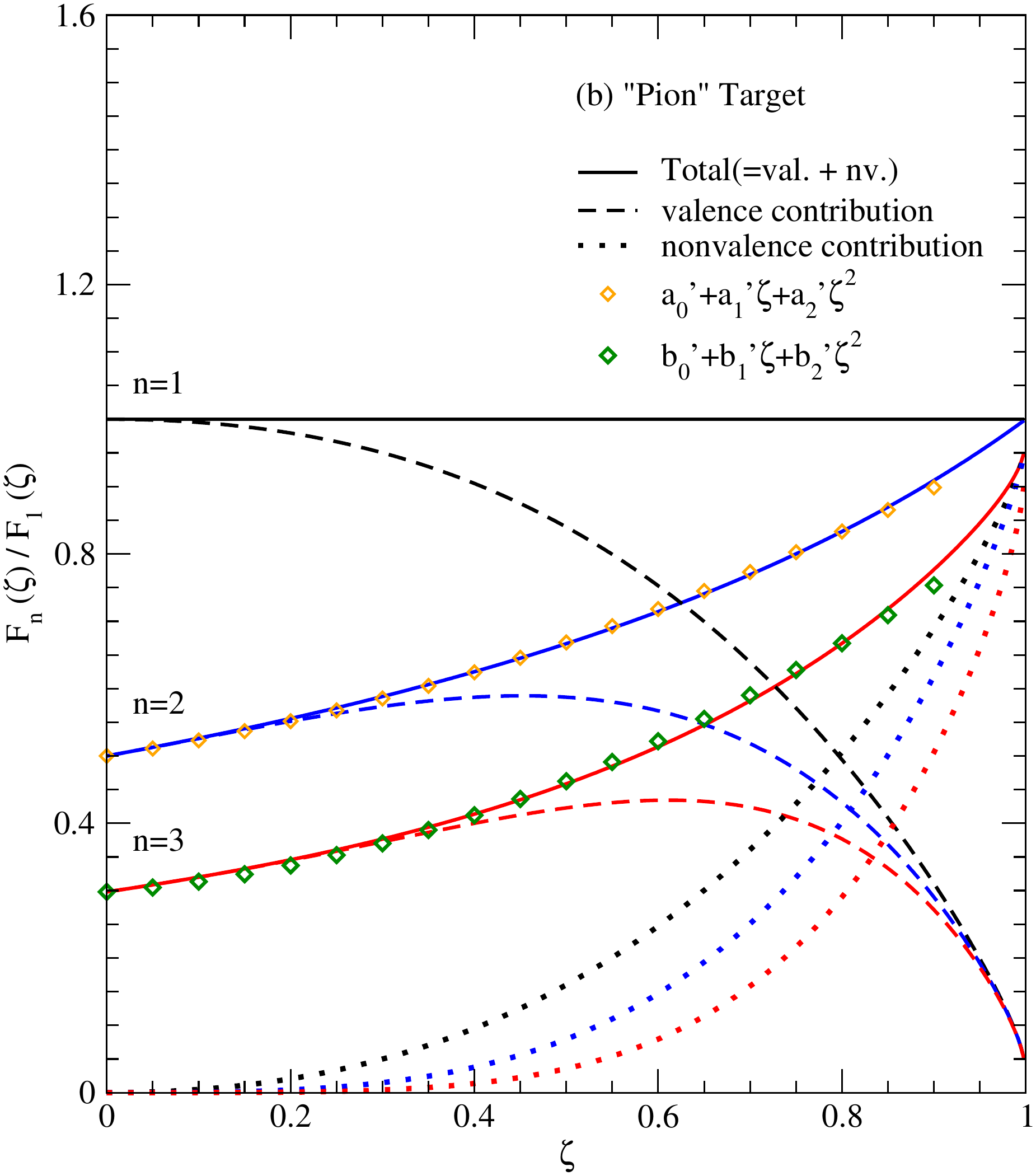}
\caption{\label{fig12}%
The first three  moments ${\bar F}_n(\zeta, t)\equiv {\bar F}_n(\zeta)$ $(n=1,2,3)$ of  $H(\zeta, x, t)$ 
given by Eq.~(\ref{eq43c}) for (a) the weakly bound ``helium'' target with $m_{Q_{1(2)}} = 2$~GeV and 
$M_T = 3.7$~GeV and 
(b) the strongly bound ``pion'' target with $m_{Q_{1(2)}}=0.25$~GeV and $M_T=0.14$~GeV, 
where $0\leq\zeta\leq 1$ corresponds to $0\leq -t\leq\infty$.}
\end{figure}

We also investigate the dependence of our results on the value of the target mass $M_T$.
In Fig.~\ref{fig12}, we show the first three moments 
${\bar F}_n(\zeta, t)\equiv {\bar F}_n(\zeta)=F_n(\zeta)/F_1(\zeta)$ $(n=1,2,3)$ of GPD $H(\zeta, x, t)$ (See Eqs.(\ref{eq43c})-(\ref{eq45c})).
We consider a weekly bound state and a strongly bound state.
Presented in Fig.~\ref{fig12}(a) are the results for the weakly bound scalar target with $m_{Q_1(2)}=2$~GeV 
and $M_T=3.7$~GeV, which is dubbed ``helium'' target.
For comparison, we also show in Fig.~\ref{fig12}(b) the results for the strongly bound scalar target with 
$m_{Q_1(2)}=0.25$~GeV and $M_T=0.14$~GeV, which is dubbed ``pion'' target~\cite{CCJO21}.
The black, blue, and red lines represent ${\bar F}_1(\zeta)$, ${\bar F}_2(\zeta)$, and ${\bar F}_3(\zeta)$, 
respectively.
The dashed, dotted, and solid lines represent the valence contributions, the nonvalence contributions, 
and their sum, respectively.
The valence and nonvalence contributions are obtained by replacing $H(\zeta, x,t)$ with 
$H_{\rm DGLAP}(\zeta, x,t)$ and $H_{\rm ERBL}(\zeta, x,t)$ in Eq.~(\ref{eq33c}), respectively.
The moments shown in Fig.~\ref{fig12} are indeed for the entire spacelike momentum transfer region since
$0\leq\zeta\leq 1$ corresponds to $0\leq -t\leq\infty$.%
\footnote{As we showed before, in ($1+1$) dimensions, $\zeta$ and $t$ are related to each other. 
But the relation depends on the mass. For example, $\zeta=0.5$ corresponds to $-t=6.85$~GeV$^2$ for 
the helium target but it corresponds to $-t=0.01$~GeV$^2$ for the pion target.}
In other words, the skewness parameter $\zeta$ is zero only at $t=0$ and the nonvalence contributions
always exist for nonzero skewness ($\zeta>0$).

Our results presented in Fig.~\ref{fig12} give the following observations.
(i) The first moments, ${\bar F}_1(\zeta)$, given by solid black lines are defined to be $\zeta$-independent 
while the sum rule for $n=1$ yields the physical EM form factor.
(ii) The redefined higher moments, ${\bar F}_{2}(\zeta)$ and ${\bar F}_{3}(\zeta)$, satisfy the 
polynomiality condition. In the figures, we plot the fitted ${\bar F}_{2}(\zeta)$ (orange diamonds) and 
${\bar F}_{3}(\zeta)$ (green diamonds) by finding the corresponding polynomials up to the second order of 
$\zeta$.
(iii) The nonvalence contribution for the weakly bound helium target does not exceed the valence contribution 
for the entire momentum transfer region as shown in Fig.~\ref{fig12}(a).
However, the nonvalence contribution for the strongly bound ``pion" is not negligible and indeed takes over 
the valence contribution at some points of $\zeta$ (or equivalently $-t$) as shown in Fig.~\ref{fig12}(b).
For example, the nonvalence contribution to the ``pion" EM form factor ($n=1$ case) is greater than the valence 
one for $\zeta \geq 0.8$ values.

\begin{figure}[t!]\centering
\includegraphics[width=0.8\columnwidth]{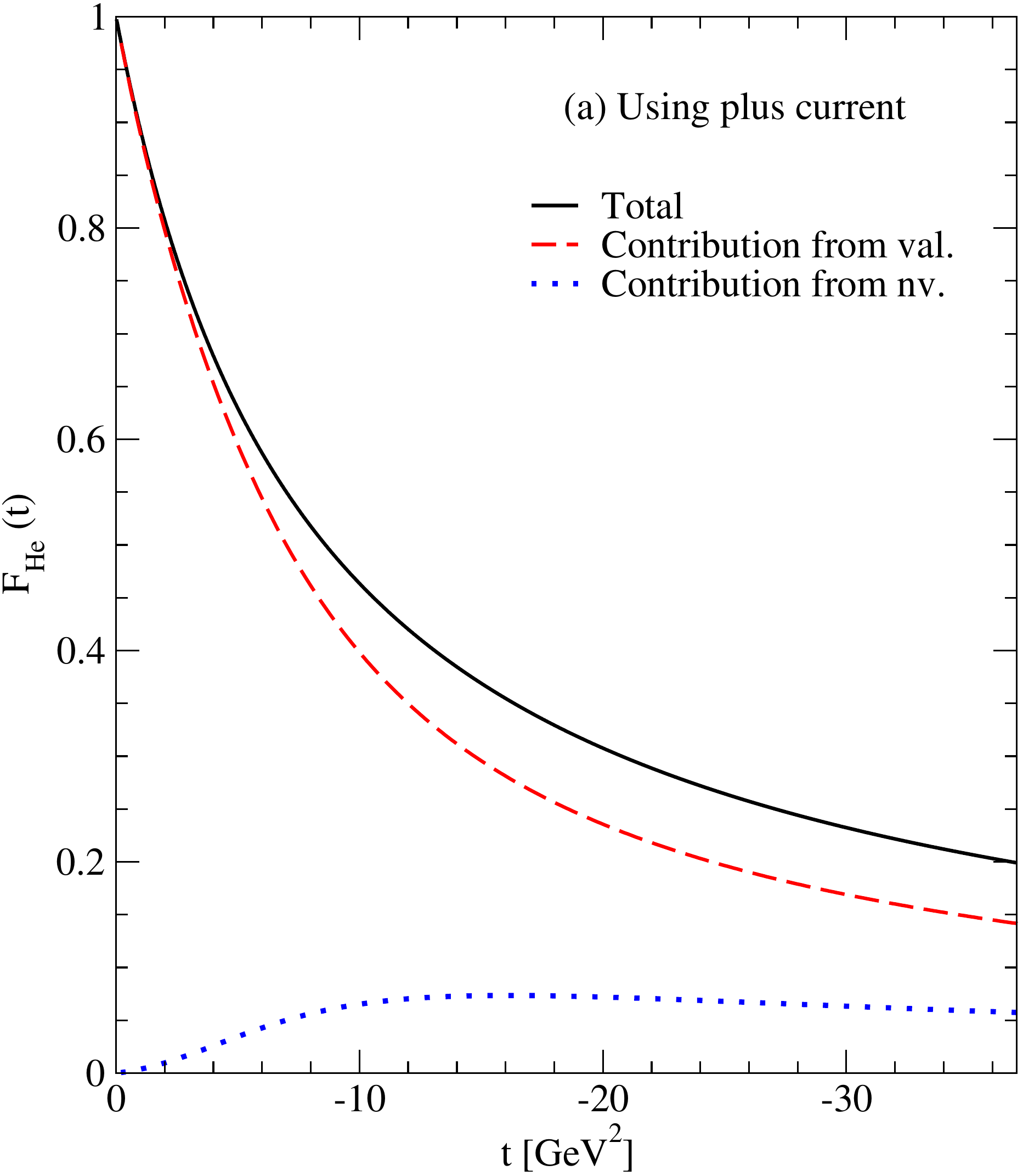}

\bigskip\bigskip

\includegraphics[width=0.8\columnwidth]{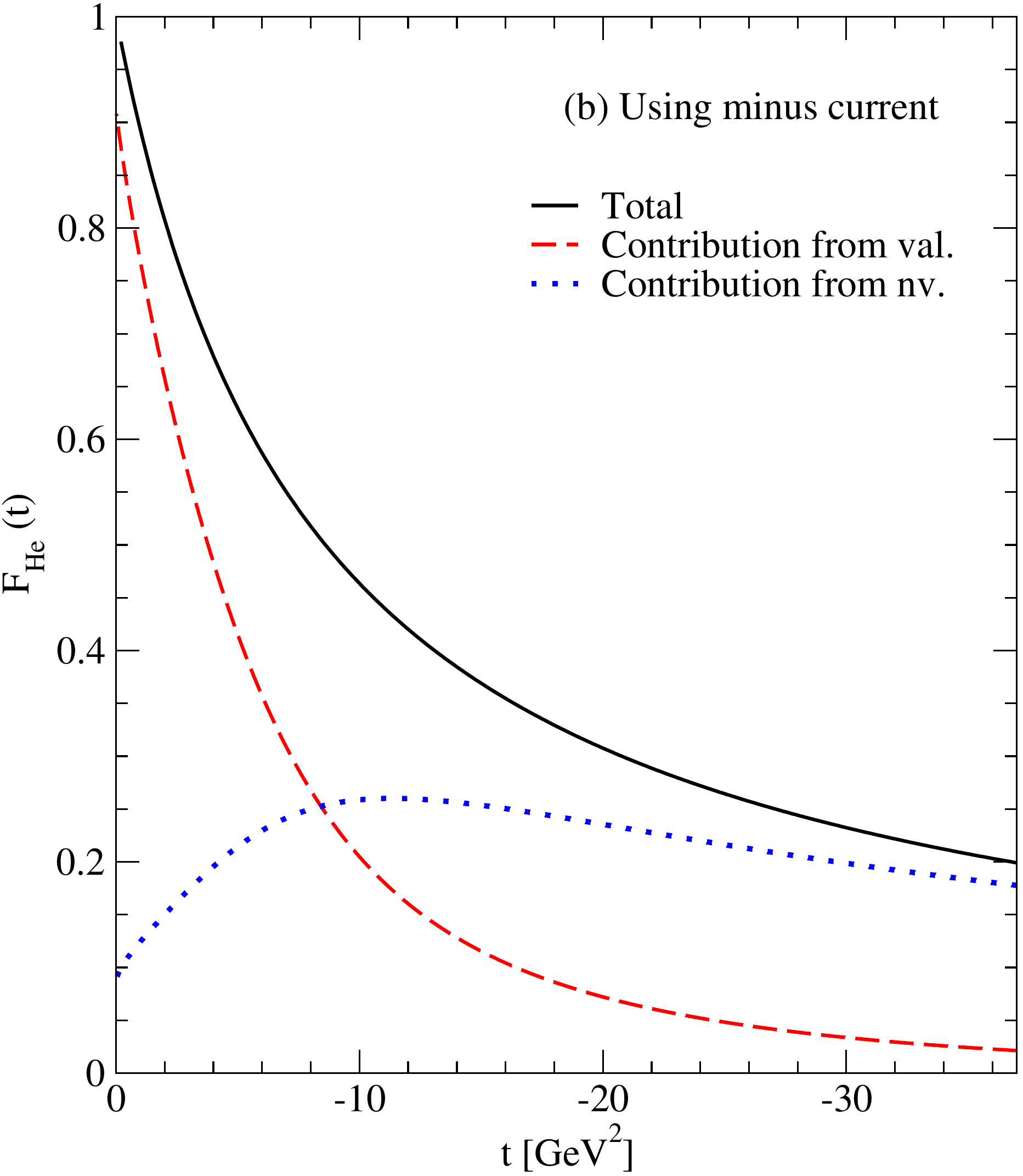}
\caption{\label{fig13}%
Electromagnetic form factor $F_{\mathcal{M}}(t)$ of the helium target as the first moment of $H(\zeta, x, t)$
obtained for $m_{Q_1(2)}=2$~GeV.
The results are obtained by using (a) the plus current and (b) the minus current.}
\end{figure}

Finally, we compute the EM form factor $F_{\mathcal{M}}(t)$ of the ``helium" target as the first moment of $H(\zeta, x, t)$ 
for the spacelike $0\leq -t < 40$ GeV$^2$ region, and the results are given in Fig.~\ref{fig13}. 
The dashed, dotted, and solid lines represent the valence contributions, the nonvalence contributions, and their 
sums, respectively. 
In particular, we obtain the form factors by taking the plus ($+$) and minus ($-$) components of the current to examine the valence and nonvalence contributions in taking different components of the current while confirming that the sum of the valence and nonvalence contributions coincide whichever component is taken. 
Our results for $F_{\mathcal{M}}(t)$ shown in Fig.~\ref{fig13}(a) are obtained by using Eqs.~(\ref{eq31c}), 
(\ref{eq32c}), and (\ref{eq44c}) together with the $``+"$ component of the current in the DVMP limit, which is exactly the 
same as our recent result reported in Ref.~\cite{CCJO21} based on the direct calculations of the triangle diagrams.
Although one typically uses the $``+"$ current to compute the form factor $F_{\mathcal{M}}(t)$, we compute here 
$F_{\mathcal{M}}(t)$ using the $``-"$ component of the current as well in our direct triangle 
diagram calculation~\cite{CCJO21}. 
For comparison, we show in Fig.~\ref{fig13}(b) the EM form factor obtained from the $``-"$  component of the current using 
Eqs.~(1) and (2) in Ref.~\cite{CCJO21}.
All these results confirm that the total EM form factors are completely the same independent of the adopted component of the 
current and either $``+"$ or $``-"$ component of the current can be used to obtain the EM form factor.
However, the decomposition of the form factor depends on which component of the current is used for the calculation.
This is apparent as shown in Figs.~\ref{fig13}(a,b), where one can see that the nonvalence contribution is quite suppressed
for the entire spacelike region when the $``+"$ current is used, 
while its contribution is not negligible but even exceeds the valence contribution for $\abs{t} \gtrsim 8.5$~GeV$^{2}$ 
when the $``-"$ current is used.
Furthermore, the nonvalence contribution for the $``-"$ current case does not vanish even at $t=0$ while it is zero for the 
$``+"$ current case.
Therefore, the decomposition of the form factor as the valence and nonvalence contributions depends on which component 
of the current is used. 
It should be interpreted with great care noting which component of the current is used.

\section{Summary and conclusion} 
\label{sec:6}

In the present work, we investigated the light-front amplitudes of the virtual meson production process off the scalar 
target in $(1+1)$ dimensions using the solvable scalar field theory. 
Noting that there is only one CFF in $(1+1)$ dimensions for this process, we obtained the analytic expressions for all 
possible LF time-ordered amplitudes shown in Fig.~\ref{fig2}. 
The obtained LF time-ordered amplitudes are individually boost-invariant, and the sum of all LF time-ordered amplitudes 
turns out gauge invariant as they must be.

With the analytic solutions of the amplitudes at hand, we investigated various quantities, including CFF, GPD, PDF, 
and EM form factor.
We first tested the ``handbag dominance'' that has been adopted in the GPD formulation for the large $Q^2$ region. 
In particular, we explored the role of the ``cat's ears'' contributions, which have been typically ignored.

To quantify the individual contribution of the LF time-ordered amplitudes, we simulated the typical mass arrangement of 
the $\gamma^{*} + \nuclide[4]{He} \to f_{0}(980) + \nuclide[4]{He}$ process. 
Our numerical results showed that the gauge invariance is largely violated if one neglects the ``cat's ears'' contribution. 
In particular, the addition of the ``cat's ears'' contribution is crucial for the low $Q^2$ region. 
Although the violation appears smaller at $Q^2 > 10$-$20$~GeV$^2$ for the imaginary part of the total amplitude, 
it is still noticeable for the real part even at large $Q^2$ as shown in Fig.~\ref{fig5}.
This appears to limit the validity of the ``handbag dominance'' to the region of small $-t/Q^2$~\cite{Ji96a,JB13}.
Our numerical calculations in $(1+1)$ dimensions presented in Fig.~\ref{fig8} show that the ``handbag dominance'' 
appears limited to the kinematic region $-t/Q^2 \lesssim 0.1$ for its applicability both to real and imaginary parts of the CFF. 
The relaxation of the condition $Q^2 \gg M^2_T$ in reaching the DVMP limit would also apply only for the forward 
production of the meson in $(3+1)$ dimensions.
Therefore, the direct use of the ``handbag dominance'' in the analyses of the proposed experiments at JLab~\cite{JLab-06},
where such small values of $-t/Q^2$ are not reached, may be treacherous necessitating great care taking into account 
the higher order $-t/Q^2$ corrections not only from the kinematic higher twist contributions but also from the dynamic 
higher twist GPDs~\cite{BMKS00,BM08,BMP12}. 
In this respect, the future Electron-Ion Collider project~\cite{EIC-21} is strongly called for the proper extraction of GPDs from
the precision experimental data off the nucleon and nuclei targets focusing on the forward angle.

In our simple $(1+1)$ dimensional model computations, the forward limit $(\zeta, t)\to 0$ of the GPD, 
$H(0,x,0) = H_{\rm DGLAP}(0,x,0)$, provides the PDF, $q_v(x)$, which could be interpreted as the probability to find 
the constituent inside the hadron as a function of the momentum fraction $x$ carried by the constituent.
It is equivalent to the square of the LF wave function, i.e., $q_v(x)= |\psi(x)|^2$. 
In $(1+1)$ dimensions, the LF spatial variable $\tilde z = x^- p^+$ is completely Lorentz-invariant providing 
an intrinsic longitudinal probability density $\varrho(\tilde{z})$. 
Our numerical results showed that the stronger bound state concentrates the density more at $\tilde z=0$ 
than the weaker bound states do, which appears to be consistent with the intuitive understanding of the bound-state system. 
The polynomiality condition for the moments of GPD appears also well satisfied.
The GPD sum rule provided the EM form factor $F_{\mathcal{M}} (t)$ confirming the valence and non-valence contributions 
that we obtained previously~\cite{CCJO21}, which corresponded to the GPD contributions from the DGLAP and 
ERBL regions, respectively.

In the calculation of the electromagnetic properties of hadrons in the LF formulation, one may use not only the
plus ($+$) component but also any other component of the current as they are supposed to give the identical results.
As shown in Fig.~\ref{fig13}, we indeed confirmed that the two components ($+$ or $-$) led to the identical form factor.
However, the decomposition of the form factor into valence and nonvalence contributions appears quite different depending
on the component of the current used in the extraction.
This indicates that it requires great care in interpreting the valence and nonvalence contributions to the form factor.

Our $(1+1)$-dimensional analyses performed in the present work would be extended to the more realistic $(3+1)$-dimensional 
analyses, where the contributions from the transverse component of the current would be important. 
In particular, the two CFFs $\mathcal{F}_1$ and $\mathcal{F}_2$ involved in the scalar meson production off the scalar target 
are independent of each other in $(3+1)$-dimensions.
Thus, the investigation of the beam spin asymmetry proportional to $\mathcal{F}_1\mathcal{F}^*_2 - \mathcal{F}_2\mathcal{F}^*_1$ 
for this process would provide a unique opportunity not only to explore the imaginary part of the hadronic amplitude in our 
general formulation but also to examine the significance of the chiral-odd GPD contribution in the leading-twist GPD formulation 
as discussed in Ref.~\cite{JCLB18}. 
The work along this line of thought is currently underway.

\medskip

\acknowledgments
The work of Y.C. and Y.O. was supported by the National Research Foundation of Korea (NRF) under Grants 
No. NRF-2020R1A2C1007597 and No. NRF-2018R1A6A1A06024970 (Basic Science Research Program).
H.-M.C. was supported by NRF under Grant No. NRF- 2020R1F1A1067990, and
C.-R.J. was supported in part by the US Department of Energy (Grant No. DE-FG02-03ER41260).
National Energy Research Scientific Computing Center supported by the Office of Science of the U.S. Department of Energy under Contract No. DE-AC02-05CH11231 is also acknowledged.

\end{document}